\RequirePackage{fix-cm}
\documentclass[smallextended]{svjour3}       
\smartqed  
\usepackage{graphicx}
\usepackage{placeins}
\usepackage{siunitx}
\usepackage{threeparttable}
\usepackage{multirow}
\usepackage{hyperref}

\newcommand*{\addthinspace}{\hskip0.16667em\relax}
\begin{document}

\title{A compact instrument for gamma-ray burst detection on a CubeSat platform II
}
\subtitle{Detailed design, assembly and validation}


\author{David Murphy \and Alexey Ulyanov \and Sheila McBreen \and Joseph Mangan \and Rachel Dunwoody \and  Maeve Doyle \and Conor O'Toole \and Joseph Thompson \and Jack Reilly \and Sarah Walsh \and Brian Shortt \and Antonio Martin-Carrillo \and Lorraine Hanlon
}


\institute{D. Murphy \and A. Ulyanov \and S. McBreen \and J. Mangan \and R. Dunwoody \and M. Doyle \and J. Reilly \and S. Walsh \and A. Martin-Carrillo \and L. Hanlon
            \at
            School of Physics \& Centre for Space Research, University College Dublin, Dublin 4, Ireland\\
            \email{david.murphy@ucd.ie}
            \and
            J. Thompson
            \at
            School of Mechanical and Materials Engineering \& Centre for Space Research, University College Dublin, Dublin 4, Ireland
            \and
            C. O'Toole
            \at
            School of Mathematics and Statistics, University College Dublin, Dublin 4, Ireland
            \and
            B. Shortt
            \at
            European Space Agency, ESTEC, 2200 AG Noordwijk, The Netherlands
}

\date{Received: date / Accepted: date}

\maketitle

\begin{abstract}
The Gamma-ray Module, GMOD, is a miniaturised novel gamma-ray detector which will be the primary scientific payload on the Educational Irish Research Satellite (EIRSAT-1) 2U CubeSat mission.  
GMOD comprises a compact (25\,mm\,$\times$\,25\,mm\,$\times$\,40\,mm) cerium bromide scintillator coupled to a tiled array of 4\,$\times$\,4 silicon photomultipliers, with front-end readout provided by the IDE3380 SIPHRA. 
This paper presents the detailed GMOD design and the accommodation of the instrument within the restrictive CubeSat form factor. The electronic and mechanical interfaces are compatible with many off-the-shelf CubeSat systems and structures. The energy response of the GMOD engineering qualification model has been determined using radioactive sources, and an energy resolution of 5.4\% at 662\,keV has been measured. 

EIRSAT-1 will perform on-board processing of GMOD data. Trigger results, including light-curves and spectra, will be incorporated into the spacecraft beacon and transmitted continuously. Inexpensive hardware can be used to decode the beacon signal, making the data accessible to a wide community.

GMOD will have scientific capability for the detection of gamma-ray bursts, in addition to the educational and technology demonstration goals of the EIRSAT-1 mission. The detailed design and measurements to date demonstrate the capability of GMOD in low Earth orbit, the scalability of the design for larger CubeSats and as an element of future large gamma-ray missions. 

\end{abstract}

\section{Introduction}
\label{intro}
\label{ch:gmod}
Astrophysical sources of gamma-rays (considered here to cover the energy range from tens of keV to several GeV) play an increasingly important role in time domain and multi-messenger astronomy \cite{MuraseBartos2019}. Gamma-ray bursts (GRBs) produce their peak energies at several hundred keV, signposting the collapse of massive stars and the mergers of neutron star binaries or neutron star-black hole binaries \cite{KumarZhang2014,MeszarosGehrels2012,GRF2009,Berger2014}. The short-lived gamma-ray transient, GRB\,170817A, was the first detected electromagnetic counterpart to a gravitational wave source \cite{abbott2017a,abbott2017b,savchenko2017,goldstein2017}. This energy range is also key to a fuller understanding of the processes operating in relativistic jet and outflow sources, nucleosynthesis occurring in novae and supernovae, the physics of compact objects, and the nature of dark matter \cite{kanbach2010,prantzos2011,Porter2011}. 

Many of the major gamma-ray satellites, including Konus-WIND \cite{Aptekar1995}, Swift \cite{Gehrels_2004}, INTEGRAL \cite{Winkler2003}, Fermi \cite{Atwood_2009} and AGILE \cite{Celesti2004}, are mature missions in their extended operations phase. The upcoming SVOM mission is designed primarily to detect, localise and follow up GRBs and other high-energy transients \cite{Atteia2020}. 
Einstein Probe is a forthcoming mission due to launch by the end of 2022 with wide-field x-ray monitoring and narrow-field x-ray follow-up capability, targeting transient astrophysical events \cite{Yuan2015,Liu2021}.
THESEUS \cite{amati2018} and the Gamow Explorer \cite{white2020} are mission concepts that focus on high redshift gamma-ray bursts and time domain multi-messenger astronomy. 
THESEUS was in contention for the fifth medium-class mission in ESA's Cosmic Vision programme but was not finally selected for implementation \cite{theseus}.
However, a space instrument covering the relatively unexplored energy band from 100\,keV to several GeV, that would be a successor to the COMPTEL instrument on the Compton Gamma-Ray Observatory in the 1990's \cite{Schonfelder1993}, has not yet been implemented, despite numerous proposals  \cite{Lebrun2003,Greiner2012,vonBallmoos2012,Tatischeff2019,deAngelis2017,Kierans2020a}. 

One of the major technical challenges to be overcome to access this relatively unexplored energy band with the required sensitivity is the need to exploit both Compton scattering and pair creation in a single instrument, which leads to significant mass and complexity in both hardware and on-board software. Fast, high light-yield scintillators \cite{vanbuis2005}, new optical detectors, such as silicon photomultipliers (SiPMs), and high-speed digital electronics, all promise to greatly improve instrument performance while reducing mass and complexity \cite{bloser2014}. 

High altitude balloons have long been used as test-beds for new space technologies. 
The performance of a compact Compton telescope design, COSI, based on Germanium \cite{kierans2017,Kierans2020b} has been demonstrated on a balloon flight, as has a Compton telescope based on SiPMs, with LaBr$_3$ \cite{bloser2016} and CeBr$_3$ \cite{Sharma2020}. 
GAGG(Ce) and Lutetium Fine Silicate (LFS) scintillators paired with an SiPM array and SIPHRA ASIC have been investigated for use in Compton telescopes \cite{nobashi2021}.
The US DoD Space Technology Programme has provided in-orbit demonstration (IOD) of the compact SIRI instrument, combining an SiPM array with an SrI$_2$ scintillator and successors SIRI-2, GARI (using GAGG scintillator) and Glowbug are in development \cite{Mitchell2019,Mitchell2021}. The technology readiness level (TRL) of these new technologies has been advanced further with the launch of the GECAM pair of micro-satellites in December 2020 which use a hemispherical array of detectors composed of LaBr$_3$ with SiPMs, to detect GRBs and other transients \cite{Zhang2019}. 

Nanosatellites (usually defined as having masses less than 10\,kg) and CubeSats (built from modular cubes, 10\,cm on a side) are also being deployed as platforms for IOD, accelerating the qualification cycle of new technologies for space. In parallel to their usefulness for IOD, the capabilities of nanosatellites for many branches of astrophysics, and for education, are of growing interest \cite{liddle2020,serjeant2020,pinilla2020,bloser2020b,Doyle2021,Walsh2020}. 
 
Improving instrument performance, while reducing development time, cost and complexity, are key enablers to ensuring the gamma-ray transient sky remains under surveillance during gravitational wave (GW) observing runs. Gamma-ray sensors on different nanosatellites in distinct orbits can provide all-sky GRB detection and localisation capability, supporting observations by larger missions and optimising the chances for discovery of electromagnetic counterparts to GW sources. Several such missions that are currently under study or being implemented include HERMES \cite{Fiore2020,Fuschino2019}, BurstCube \cite{Perkins2020,Smith2019} and GRBAlpha \cite{Pal2020} for CAMELOT \cite{Ohno2020}.
GRID is a student-led GRB CubeSat network, the first of which was launched in 2018 \cite{wen2019}. The GRID GRB detector is based on GAGG scintillator with SiPMs. IGOSAT is a student gamma-ray 3U CubeSat based on CeBr$_3$, scheduled for launch in late 2022 \cite{Phan2018}. 

This paper describes the CubeSat GRB detector, GMOD, and its incorporation as the main science payload in EIRSAT-1, the ``Educational Irish Research Satellite'' which is planned to be Ireland's first satellite \cite{Murphy2018}. GMOD is a novel CubeSat payload in its use of CeBr$_3$ with SiPMs. 
Section~\ref{sec:gmodforeirsat} briefly outlines the evolution that has led to the current GMOD design. Sections~\ref{sec:gmoddetector} and \ref{sec:gmodmotherboard} present the configuration of the detector assembly and the motherboard in detail. Section~\ref{sec:gmoddata} gives a comprehensive overview of both the TTE data generated by GMOD, and the light-curves, spectra and triggers generated by the on-board computer (OBC). The results of the laboratory characterisation of GMOD are presented in Section~\ref{sec:gmodchar} and these results are compared to the performance of other CubeSat payloads in Section~\ref{sec:gmoddiscussion}.

A companion paper \cite{Murphy2021} presents the simulation study for GMOD, including predicted effective area as a function of energy and off-axis angle, and estimated GRB detection rates.
The environmental qualification campaign is described in \cite{Mangan2021}.

%

\section{GMOD for EIRSAT-1}
\label{sec:gmodforeirsat}
GMOD evolved from a prototype building block module for the pixellated calorimeter of a full Compton-pair telescope. This gamma-ray detector (GRD) \cite{ulyanov2016} used a monolithic LaBr$_3$(Ce) crystal scintillator with an SiPM array, read out by a low--power ASIC called `SIPHRA' \cite{Meier2016} (Section~\ref{sec:gmodasic}). The scintillator was subsequently replaced by CeBr$_3$, due to its lower intrinsic background \cite{ulyanov2017} and the GRD was flown on a high-altitude balloon \cite{murphy2021b}. 

\begin{figure}
\centering
\includegraphics[height=7.4cm]{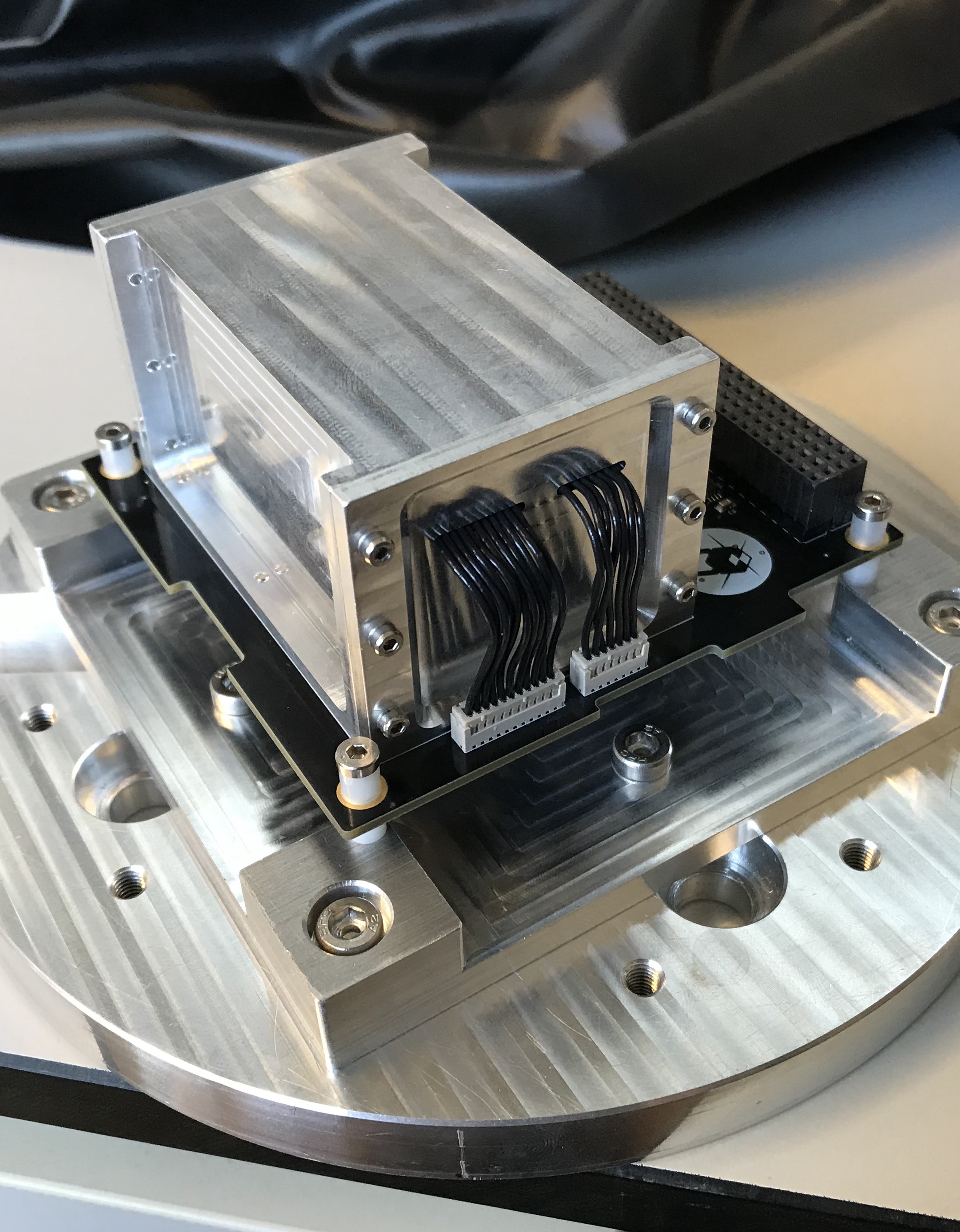}
\hspace{0.2cm}
\includegraphics[height=7.4cm]{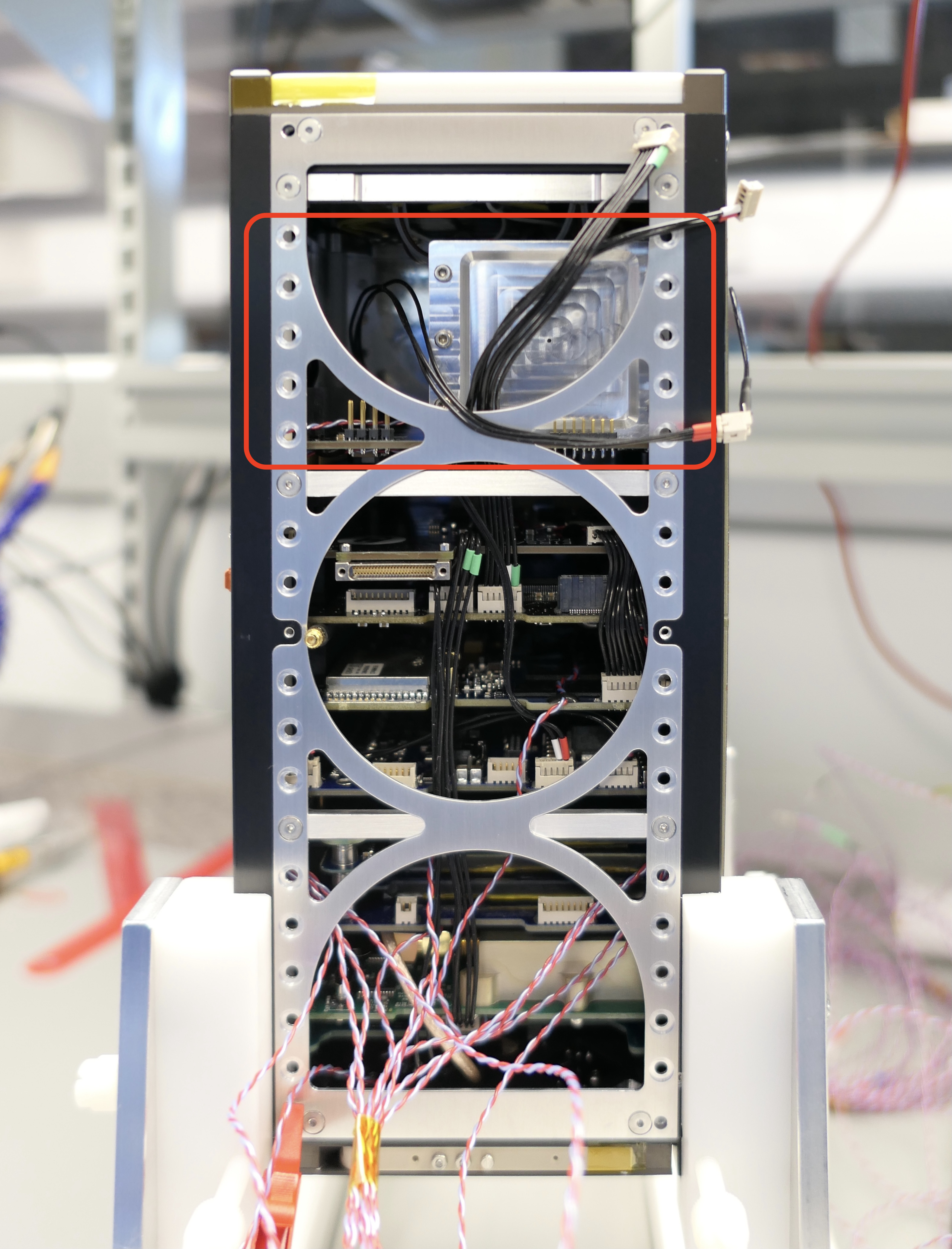}
\caption{GMOD and its accommodation within the EIRSAT-1 stack. \emph{Left:} The GMOD environmental qualification model (EQM) comprising the aluminium-enclosed detector assembly and the black motherboard PCB mounted on a test fixture. \emph{Right:} GMOD in the EIRSAT-1 EQM. GMOD is in the aluminium housing near the top of the stack, indicated by the red rectangle.}
\label{fig:gmodeirsat1}
\end{figure}

In early 2017, the initial design for GMOD as a CubeSat payload was developed as part of the student-led EIRSAT-1 mission concept \cite{Murphy2018}, with much of the architecture inherited from the GRD. The discovery of GW170817/GRB 170817A coincided with the approval of EIRSAT-1 by ESA's educational Fly Your Satellite! programme and motivated the additional requirement for GRB detection. 
Figure~\ref{fig:gmodeirsat1} shows GMOD and its accommodation within the EIRSAT-1 2U stack. 

\begin{figure}
\centering
\includegraphics[width=\textwidth]{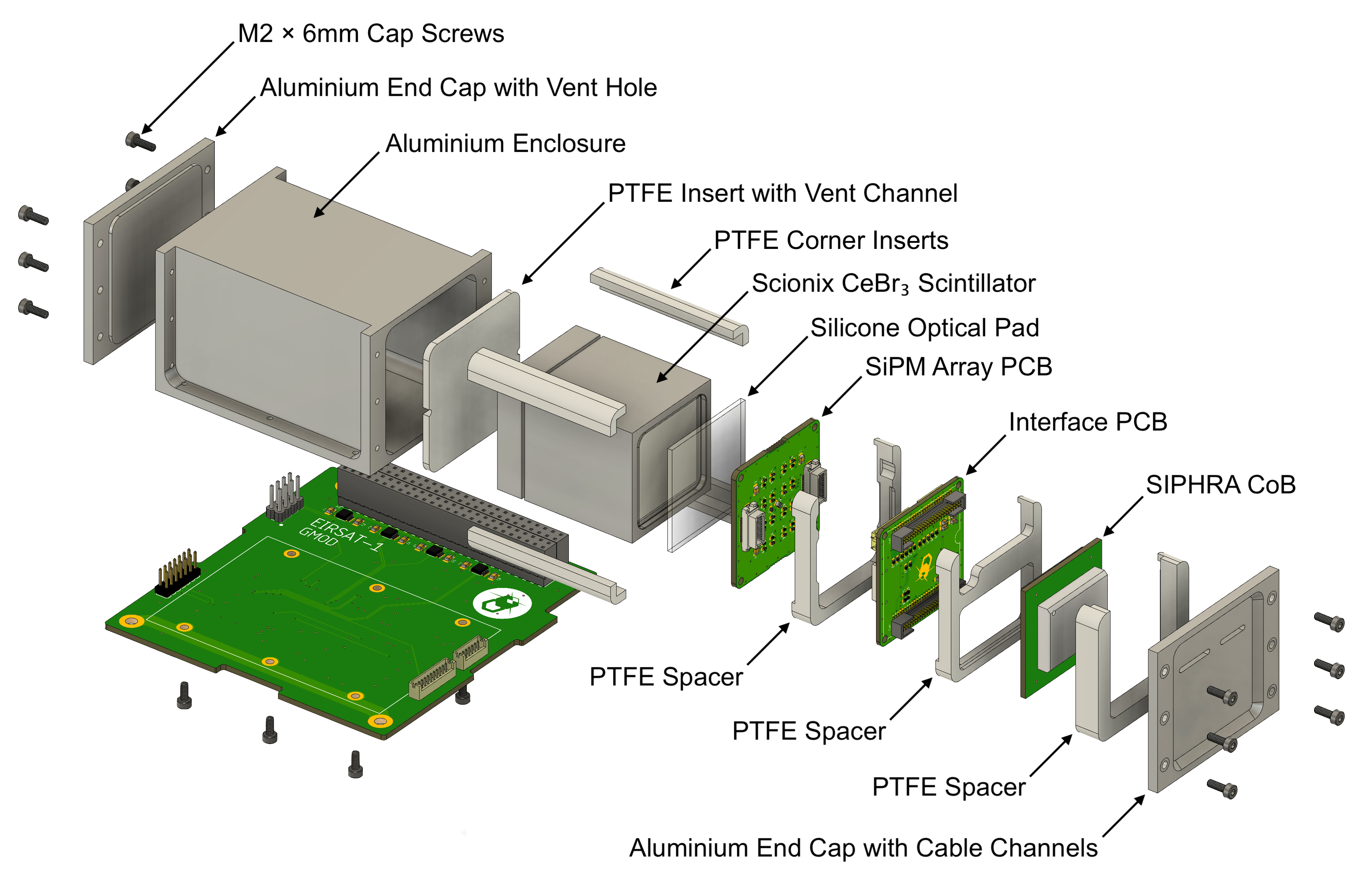}
\caption{Exploded view showing the GMOD detector assembly above, and the CubeSat-compatible GMOD motherboard beneath.}
\label{fig:explodedview}
\end{figure}

The detector assembly (Sections~\ref{sec:gmoddetector}) and the motherboard (Section~\ref{sec:gmodmotherboard}) are the two main elements of GMOD and are shown in the exploded view in Figure~\ref{fig:explodedview}. 

\section{Detector Assembly}
\label{sec:gmoddetector}
\begin{figure}
\includegraphics[width=0.8\textwidth]{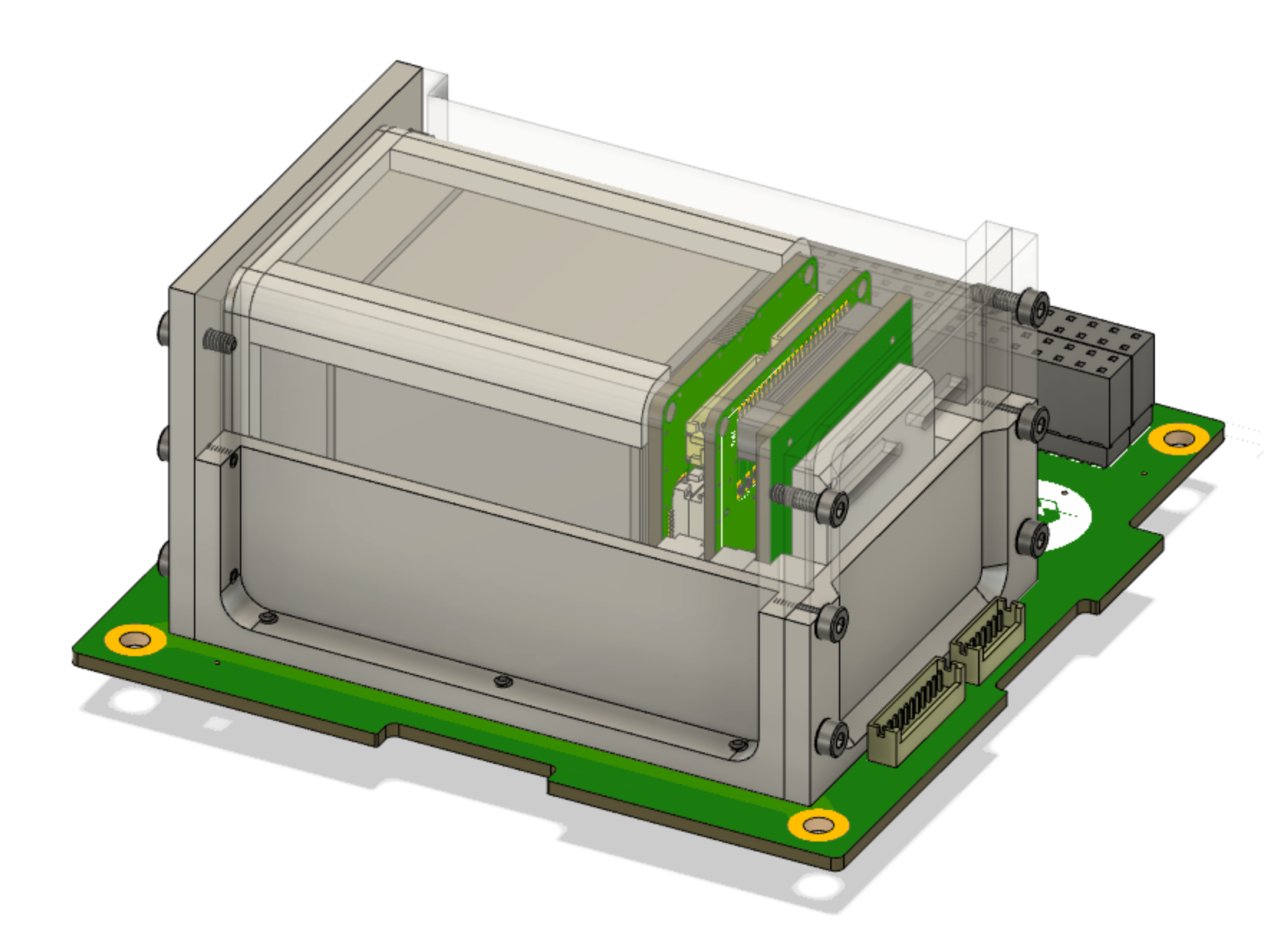}
\caption{Assembled configuration of GMOD showing the detector assembly attached to the motherboard PCB. The light-tight enclosure of the detector assembly has been rendered partially transparent to reveal the internal components. Inside the enclosure, the electronics stack sub-assembly can be seen to the right of the scintillator. The detector assembly measures 75\,mm\addthinspace$\times$\,51\,mm\addthinspace$\times$\,42\,mm.}
\label{fig:transparenttop}
\end{figure}

As shown in Figure~\ref{fig:explodedview}, the GMOD detector assembly consists primarily of a scintillator, SiPM optical detectors, and the SIPHRA which is used to process and digitise the analog signals from the SiPMs, all contained within a light-tight enclosure.
One of the advantages of using an ASIC like SIPHRA is that the readout electronics can be made sufficiently small to be included in the detector assembly (Figure~\ref{fig:gmodeirsat1}), eliminating bulky harnesses to carry analog SiPM signals to the motherboard.
Within the enclosure, the instrument is divided into two major sub-assemblies, the scintillator and an electronic stack.
The latter comprises the SiPM array, SIPHRA, and an interface PCB which ties all of the electronics in the stack together and provides the harnessed interface to GMOD's motherboard (Section~\ref{sec:gmodmotherboard}).
Figure~\ref{fig:transparenttop} shows these two sub-assemblies inside the enclosure.

%

\subsection{Scintillator}
GMOD uses a 25\,$\times$\,25\,$\times$\,40\,mm\,$^{\rm 3}$ cerium bromide (CeBr$_3$) crystal scintillator supplied by Scionix enclosed within a hermetically sealed unit (Figure~\ref{fig:sipmarrays}). 
The hermetic enclosure has external dimensions of 33\,$\times$\,33\,$\times$\,43.8\,mm\,$^{\rm 3}$ and is made of aluminium with a quartz window to allow the scintillation light to exit.
The 2\,mm thick quartz window is placed on one of the square 33\,mm\,$\times$\,33\,mm faces of the enclosure, exposing a 25\,mm\,$\times$\,25\,mm face of the CeBr$_3$ crystal.
The remaining five sides are wrapped with a PTFE light reflector.

The CeBr$_3$ scintillator was selected for a number of reasons. CeBr$_3$ has similar performance characteristics to LaBr$_3$ \cite{Billnert2011} but is available at a much lower cost. CeBr$_3$ does not suffer from the high intrinsic radiation of LaBr$_3$ caused by decay of $^{138}$La.
The smaller 25\,mm square cross-section of the CeBr$_3$ crystal is well matched to the J-series SiPM array, which has an active area extent of 25.06\,mm\,$\times$\,25.06\,mm.


\subsection{SiPM Array}
To measure the scintillation light produced by the crystal, GMOD uses sixteen J-series 60035 SiPMs  produced by SensL (now ON Semiconductor). The J-series has a microcell fill factor of 75\% and 22,292 microcells for the 6\,mm variants and a dark count rate of 50\,kHz/mm$^2$. The J-series SiPM is available in a through-silicon via (TSV) package with a Ball Grid Array (BGA) interface. With a 35\,\si{\micro\metre} microcell size, the active surface area of the J-series SiPM spans 6.07\,mm\,$\times$\,6.07\,mm and is accommodated in the 6.13\,mm\addthinspace$\times$\,6.13\,mm TSV package.

\begin{figure}
\centering
\includegraphics[height=6.8cm]{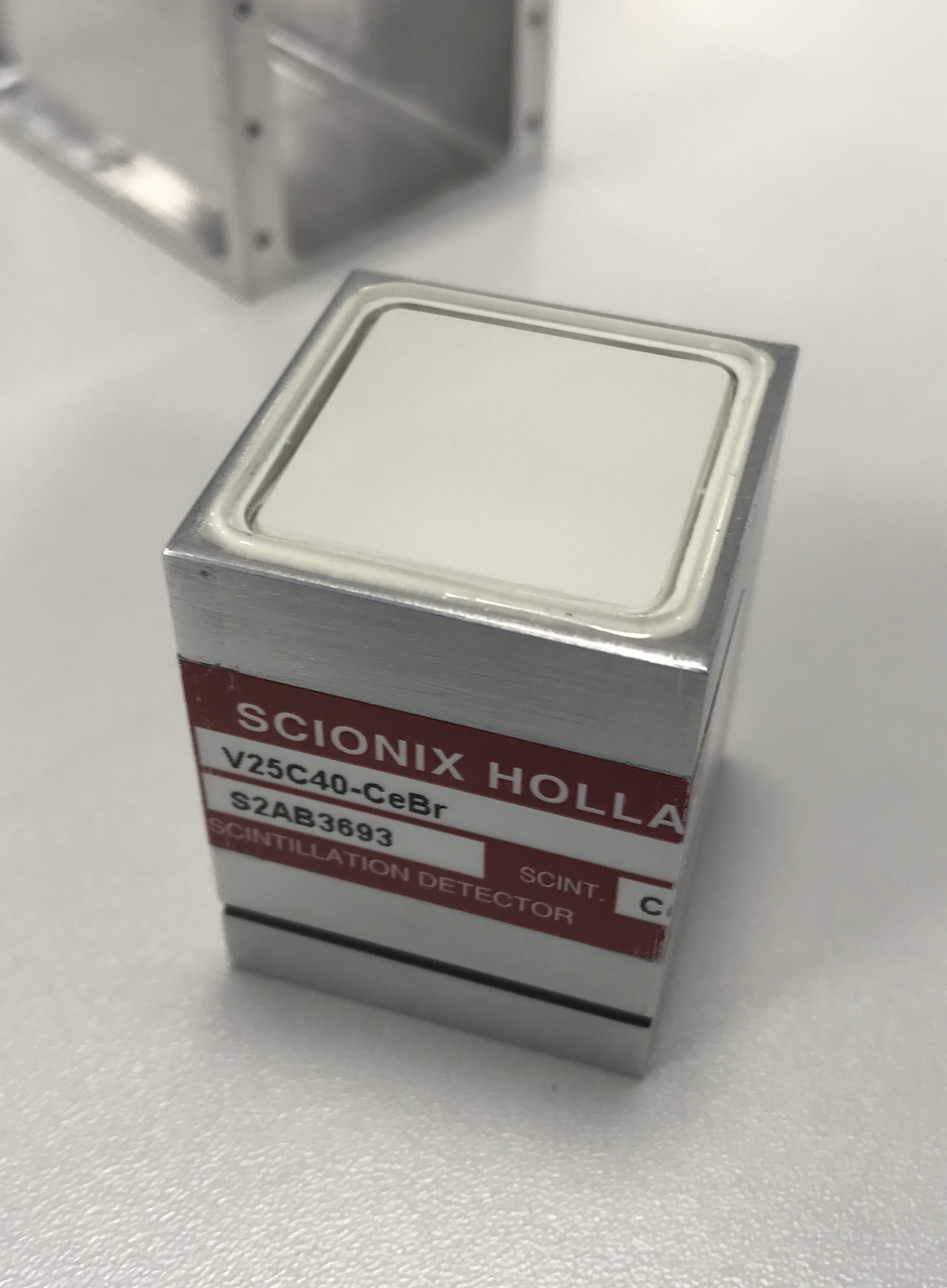}
\hspace{0.3cm}
\includegraphics[height=6.8cm]{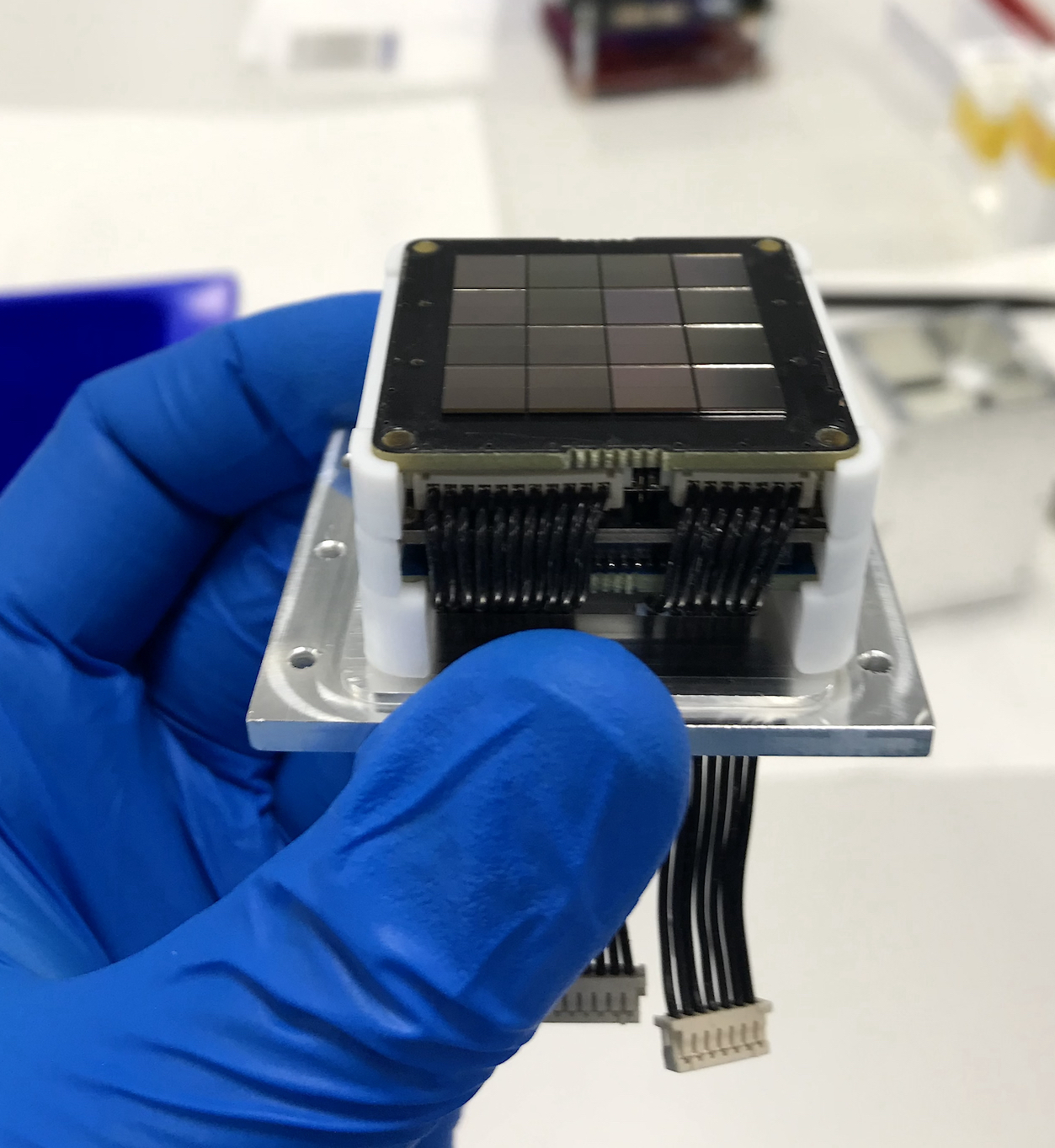}
\caption{\emph{Left:} The CeBr$_3$ crystal scintillator. \emph{Right:} The J-series 4\,$\times$\,4 SiPM array pictured atop the detector assembly electronics stack. The interface PCB and SIPHRA ASIC can be seen below the array separated by a series of supporting PTFE spacers.}
\label{fig:sipmarrays}
\end{figure}

The SiPMs are arranged on a custom 35\,mm\,$\times$\,35\,mm PCB using an automated pick-and-place process to produce a 4\,$\times$\,4 array with a nominal 200\,\si{\micro\metre} spacing, matching the recommended value (Figure~\ref{fig:sipmarrays}).
This gives a total tiled SiPM span of 25.12\,mm\,$\times$\,25.12\,mm while the active microcells within the full array have an extent of 25.06\,mm\,$\times$\,25.06\,mm.

The SiPM array is arranged in a common-anode configuration, with all SiPM anodes being connected to a common negative bias supply via independent low-pass filters consisting of a 51\,$\Omega$ resistor and a 10\,nF capacitor.
The filters are contained on the reverse side of the PCB which also has two Hirose DF17 board-to-board connectors which provide the interface to the array.
The cathode of each SiPM is connected directly to the ASIC inputs via the board-to-board connectors.
The pinout of the board-to-board connections was designed to be symmetrical so that it is permissible to attach the SiPM array upside-down, removing a potential assembly error.

SiPM performance is affected by temperature, notably SiPM breakdown voltage varies with temperature \cite{jseries}.
As SiPM gain is a function of the over-voltage above breakdown at which the SiPMs are biased, this temperature dependence of the breakdown voltage leads to an overall temperature dependence for the instrument calibration.
The reverse side of the SiPM array therefore contains a PT100 temperature sensor placed at the centre of the board to measure the SiPM temperature.
This sensor is digitised by the SIPHRA ASIC (Section~\ref{sec:gmodasic}) each time the SiPM outputs are sampled allowing for periodic adjustments to be made to the bias voltage (Section~\ref{sec:gmodbiaspsu}) or for appropriate temperature-dependent calibration to be applied on the ground.
While no active thermal control is used to regulate the SiPM temperature, a passive system is used to insulate the arrays from external temperature fluctuations.
as discussed in Section~\ref{sec:enclosure}).

\begin{figure}
  \includegraphics[width=0.8\textwidth]{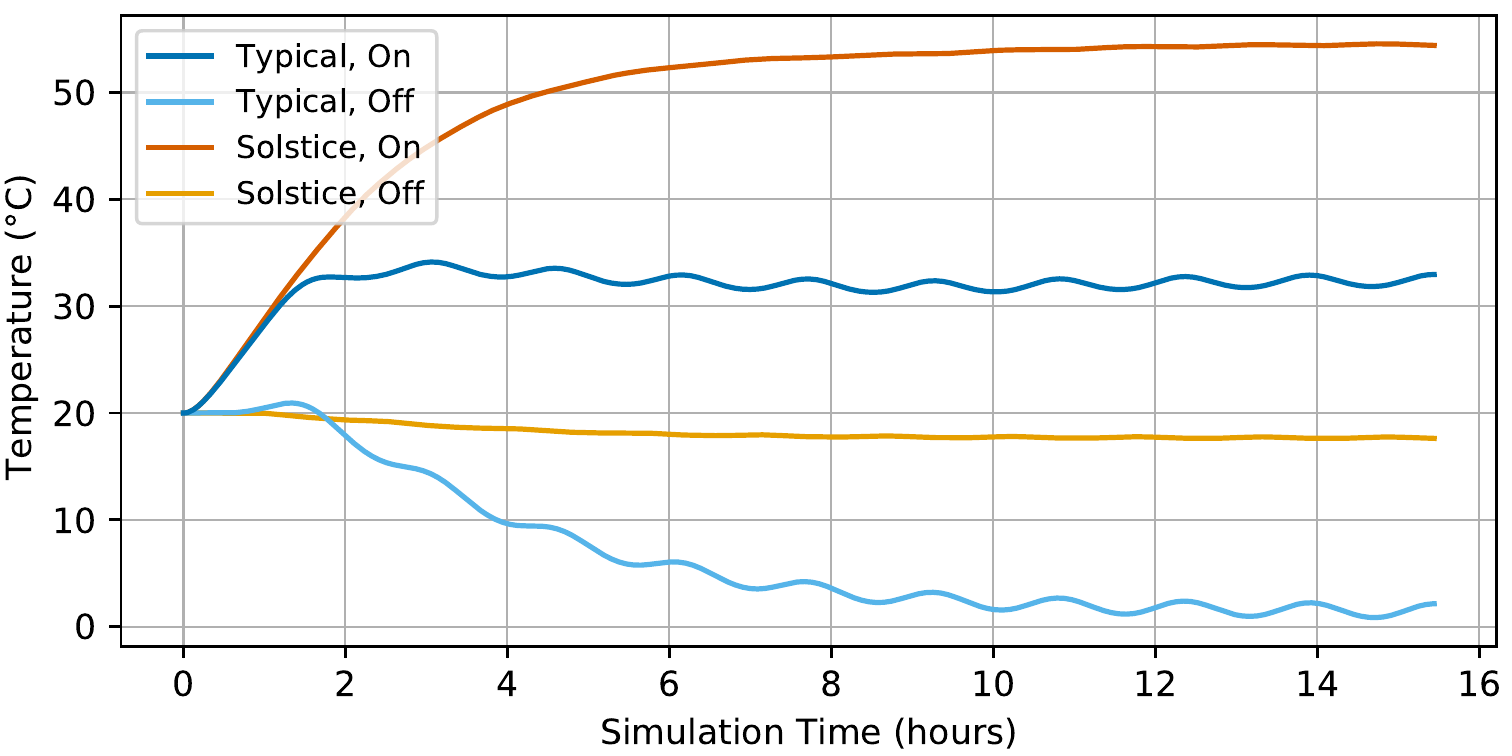}
  \caption{The temperature of the SiPM array as predicted by the EIRSAT-1 thermal analysis simulation. The typical and solstice cases for an ISS-like orbit are shown, both with (On) and without (Off) the effects of power dissipation within the spacecraft. Note: the values shown here carry uncertainties of $\pm$10$^{\circ}$C \cite{ECSS_E_HB_31_03A_2016}. }
  \label{fig:sipmtemps}
\end{figure}

As part of the EIRSAT-1 critical design review, a thermal analysis of the spacecraft was performed considering a number of scenarios for the planned ISS-like orbit.
The predicted SiPM temperatures for the planned ISS-like low Earth orbit are shown in Figure~\ref{fig:sipmtemps}.
The initial conditions for the simulation set the spacecraft temperature to 20$^{\circ}$C and for a typical orbit, a quasi-equilibrium SiPM temperature is reached between $\sim$31$^{\circ}$C and $\sim$34$^{\circ}$C.
For an ISS-like orbit, during the solstices, it is possible for the spacecraft to spend several days in un-eclipsed sunlight.
During this time the spacecraft temperature increases, pushing the SiPM temperature to approximately 55$^{\circ}$C, though with lower variability due to the lack of thermal cycling.
It should be noted that due to the approximations involved in thermal modelling these values carry an uncertainty of $\pm$10$^{\circ}$C \cite{ECSS_E_HB_31_03A_2016}.

\begin{figure}
\centering
\includegraphics[height=3.6cm]{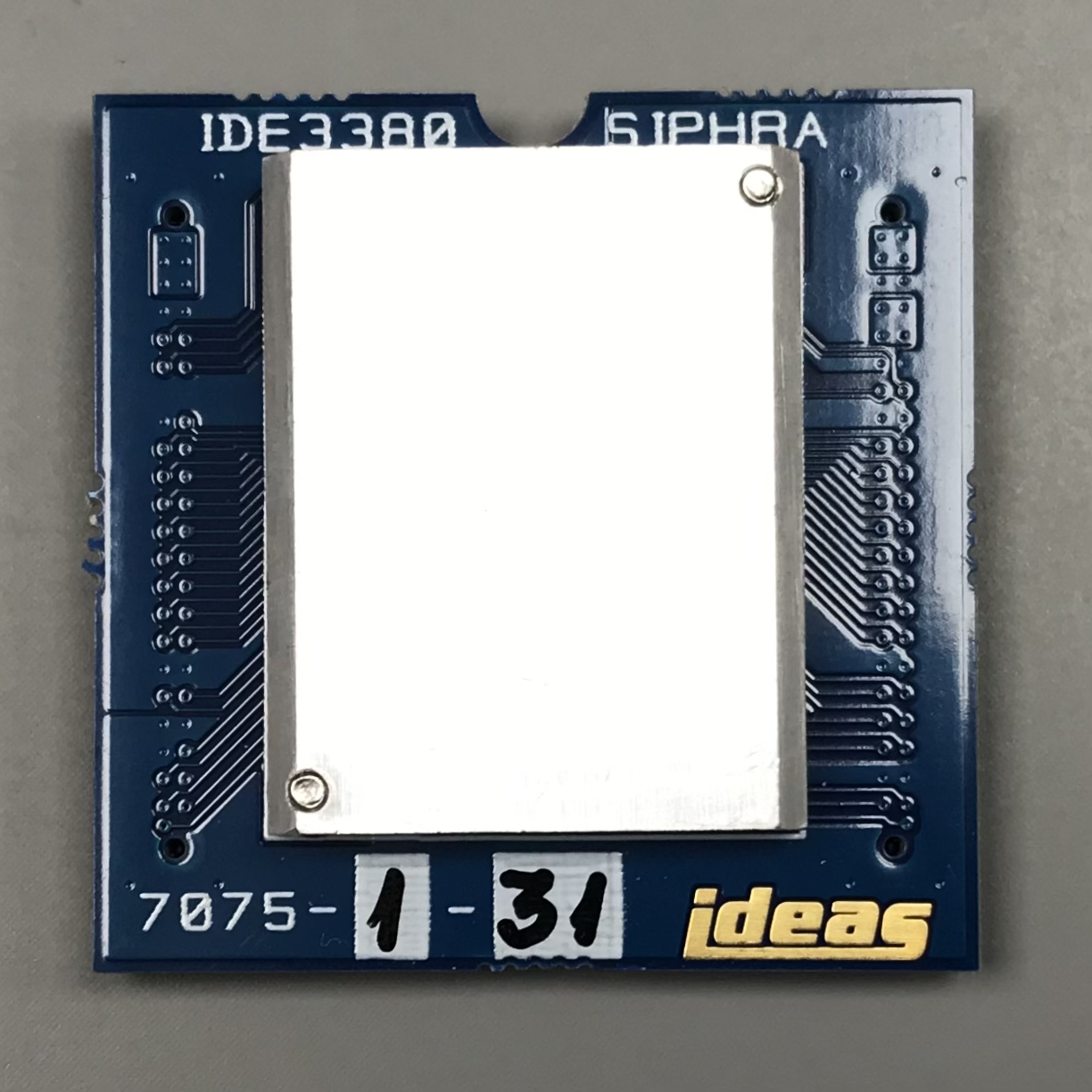}
\hspace{0.3cm}
\includegraphics[height=3.6cm]{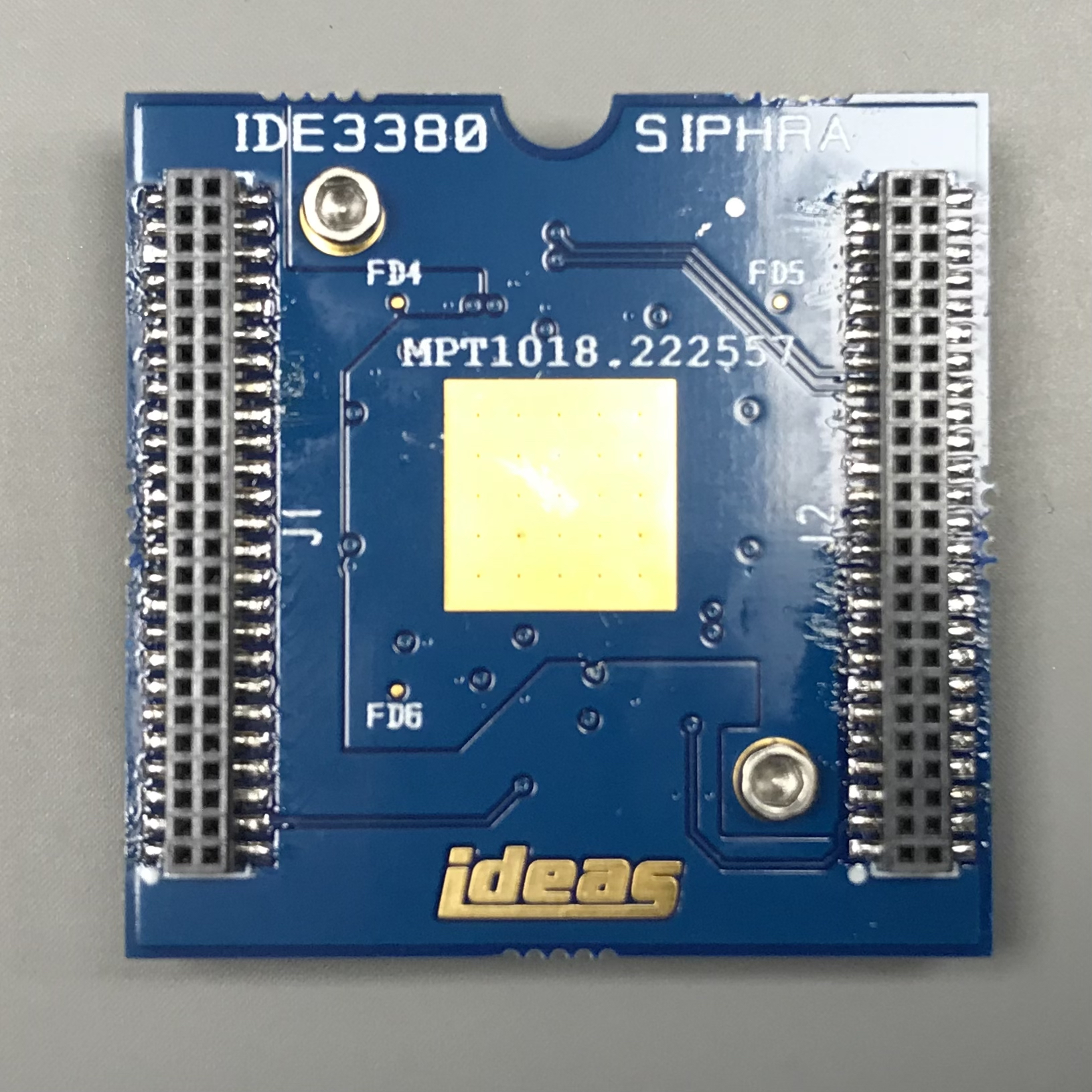}
\hspace{0.3cm}
\includegraphics[height=3.6cm]{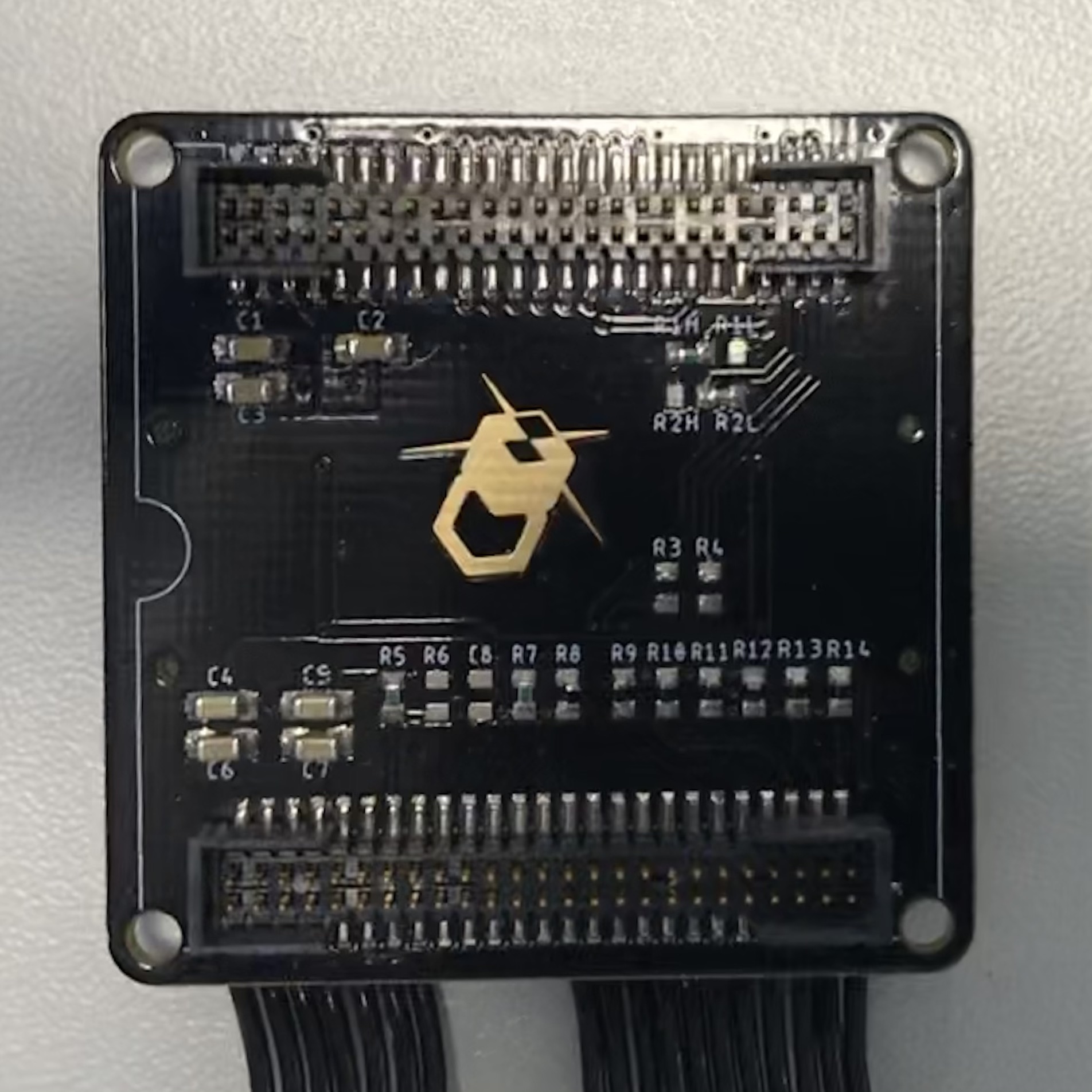}
\caption{\emph{Left:} SIPHRA CoB top view showing the aluminium chip cover which protects the silicon die. \emph{Centre:} SIPHRA CoB underside. \emph{Right:} Interface PCB. This side interfaces to the SIPHRA CoB while the reverse side interfaces to the SiPM array and has the harness connectors visible in Figure~\ref{fig:sipmarrays}. }
\label{fig:siphra}
\end{figure}

%

\subsection{SIPHRA}
\label{sec:gmodasic}
The SIPHRA (Silicon Photomultiplier Readout ASIC) \cite{ide3380} is used to read out and digitise the analog pulses from the SiPM array (Figure~\ref{fig:siphra}).
While SIPHRA was developed by Integrated Detector Electronics AS (IDEAS) as a general purpose readout IC for photon detectors, it was designed using the GRD's SiPM array as a reference. As GMOD's array topology is identical to that of the GRD, SIPHRA is therefore particularly well suited to this application. 
SIPHRA was designed with with low-power considerations and with radiation protections, including latch-up immunity, single event upset mitigation and error correction which have been verified by radiation hardness testing \cite{Stein2019}.

SIPHRA has sixteen channels for SiPMs and an internal seventeenth summing channel which provides a sum of the sixteen inputs. The SIPHRA analog front-end is described in Section~\ref{sec:frontend}. Control and configuration of SIPHRA (Section~\ref{sec:control}) is achieved through a series of registers accessed by a Serial Peripheral Interface (SPI).

SIPHRA operates at 3.3\,V and has three power nets (digital, analog, and I/O) which are supplied by separate low-drop-out voltage regulators on the GMOD motherboard. SIPHRA requires an external 80\,\si{\micro\ampere} constant reference current to generate internal biases. This reference current is generated by a constant-current source on the GMOD motherboard (Section~\ref{sec:gmodmotherboard}).

\subsubsection{SiPM Analog Frontend}
\label{sec:frontend}
As GMOD uses a common-anode SiPM array, it can utilise SIPHRA's current mode input stages (CMIS) which work with negative charge and allow for a range of input attenuation from $-$16\,nC to $-$0.4\,nC. 
Each channel's CMIS is connected to a current comparator and to a charge comparator and pulse height spectrometer via a current integrator.
This setup is replicated for the summing channel, which is connected to all sixteen CMIS.
The trigger thresholds of the comparators are configured using a series of 8-bit DACs and a configurable hold unit allows any or all of the comparators to trigger readout of the spectrometer as well as indicating the trigger type and channel for each readout.

The pulse height spectrometer consists of a pulse shaper, a track \& hold circuit, and an output multiplexer to SIPHRA's ADC.
When a trigger occurs, a hold is asserted on all channels allowing the ADC to digitise each channel in turn.
The same ADC is used to digitise the SiPM array temperature with each trigger using the PT100 mounted on the reverse of the array.

\subsubsection{Control and Data Interfaces}
\label{sec:control}
Trigger data from SIPHRA is output using a serial interface which operates at the 1\,MHz input clock rate. 
The serial output is controlled by a readout controller which waits for a hold that is either generated internally by any of the enabled comparators, externally via SPI command, or with the dedicated hold input  connected to the GMOD motherboard. The controller then digitises enabled channels in sequence.
The digitised pulse height values are combined with a series of trigger flags and channel IDs, serialised, and transmitted to the GMOD motherboard from the serial output.

Various logic-level inputs and outputs are provided for direct interfacing to certain features.
GMOD utilises the external hold input and the error output.
The external hold input is used to generate external triggers for baseline noise determination.
The error output indicates if any of SIPHRA's configuration register parity checks have an error which signals an error in the registers potentially caused by a single event upset.
 
\subsection{Interface PCB Design}
\label{sec:gmodasicpcb}

\sloppy
The interface PCB (Figure~\ref{fig:siphra}) uses the Chip-on-Board (CoB) version of SIPHRA which is wire-bonded to a PCB, but is contained on a 32\,mm\addthinspace$\times$\,32\,mm PCB produced by IDEAS with two 50-pin header connectors. SIPHRA CoB is provided as a socketed component requiring no difficult processes, such as wire-bonding, to integrate it with the custom electronics. Although this version of SIPHRA requires more volume than other versions (e.g. BGA), making its accommodation within the GMOD detector assembly more challenging, it significantly reduces assembly complexity. 

The detector assembly interface PCB therefore acts as a three-way interface between the SiPM array, the SIPHRA CoB package, and the GMOD motherboard. The PCB measures 35\,mm\,$\times$\,35\,mm to match the size of the SiPM array PCB and couples directly to the rear of the array using Hirose DF17 board-to-board connectors matching those on the array.
The reverse side features two Samtec FTM board-to-board to match the SIPHRA CoB interface, resulting in a small detector assembly electronics stack of three PCBs, as shown in Figure~\ref{fig:explodedview}.

The connection to the GMOD motherboard is achieved through two small harnesses. An 11-pin connection is used to carry digital signals while a 7-pin connection is used for power connections. Both harnesses use Hirose DF13 connections which are extensively used in the EIRSAT-1 spacecraft.
The digital connection carries the following: two ground lines; the external hold input, which externally triggers ADC readout; the error output signal,  which indicates a parity error in the configuration registers; reset; SPI configuration interface; the 1\,MHz system clock input and the 1\,MHz serial data output.
The power connection carries two ground lines, the 80\,\si{\micro\ampere} constant reference current, 3.3\,V supply for the three SIPHRA supply nets, and the SiPM negative bias voltage.

Various passive electronic components required to support SIPHRA, such as decoupling capacitors, pull-up resistors and input protection resistors, are also included on the interface PCB.

\subsection{Enclosure} 
\label{sec:enclosure}
The GMOD detector assembly is contained within an aluminium light-tight enclosure which physically supports the assembly components and prevents any external light from reaching the SiPMs. The enclosure has external dimensions of 75\,mm\addthinspace$\times$\,51\,mm\addthinspace$\times$\,42\,mm and is mounted directly to the GMOD motherboard as shown in Figures~\ref{fig:explodedview} \& \ref{fig:transparenttop}.

The design of the enclosure takes account of the requirement to protect the delicate scintillator crystal, particularly during launch. 
For strength, the GMOD enclosure is made from Aluminium 6082, the same material as the CubeSat structure. The enclosure is essentially a square-tube-like structure (Figure~\ref{fig:gmodeirsat1}) with flanges to allow the attachment of two end-caps and to allow mounting of the detector assembly to the motherboard.
The two openings are necessary to achieve a good surface finish along the full length of the bore by machining from both ends of the enclosure while  also providing a convenient disassembly mechanism as the detector assembly components can be pushed clear through the enclosure.
The internal clear cross section of the enclosure has dimensions of 38\,mm\,$\times$\,38\,mm with 3.5\,mm diameter rounded corners.

\begin{figure}
\includegraphics[height=4.85cm]{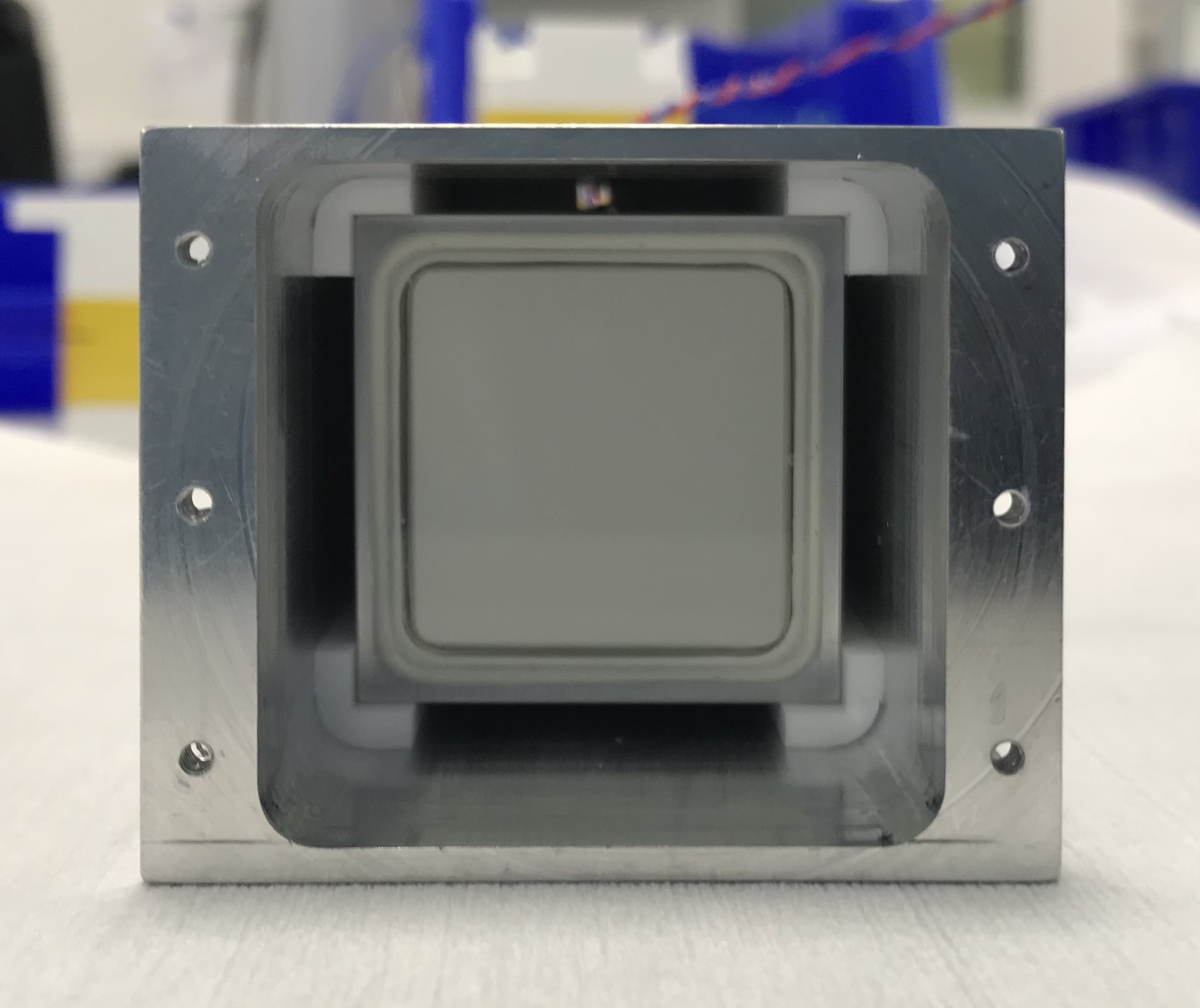}
\hspace{0.3cm}
\includegraphics[height=4.85cm]{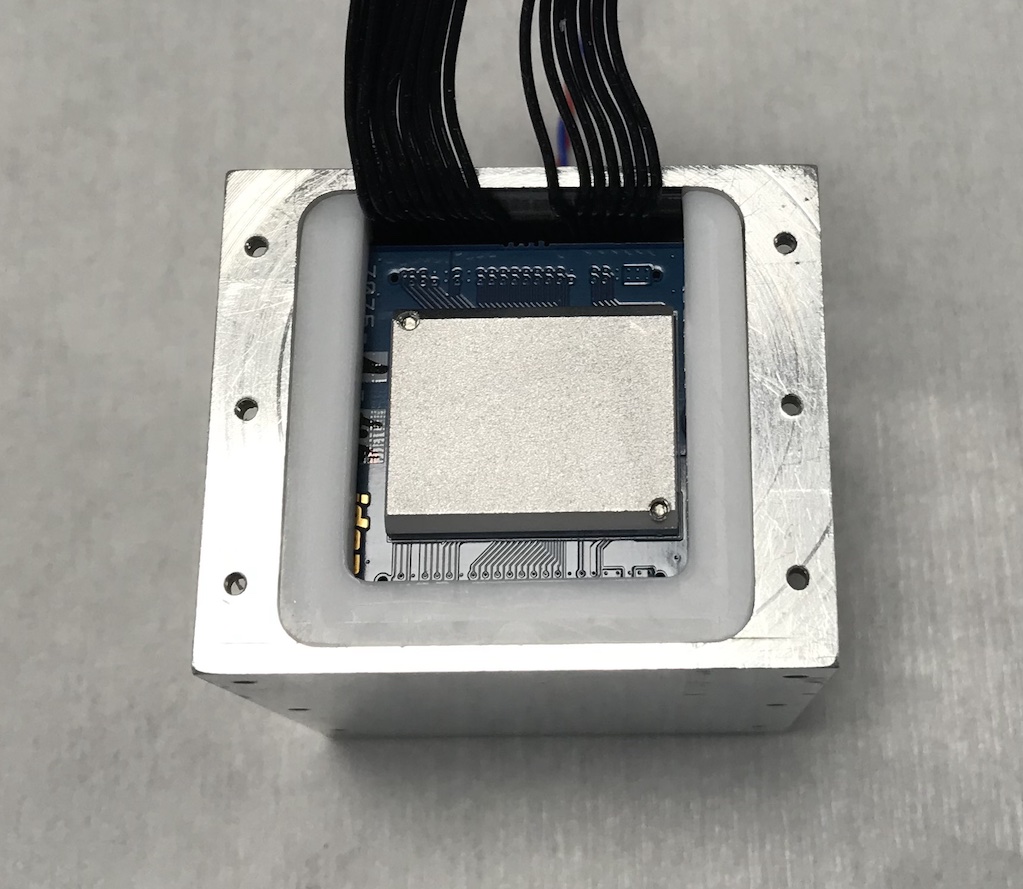}
\caption{\emph{Left:} The CeBr$_3$ crystal scintillator held securely within the aluminium enclosure by the four PTFE corner spacers. \emph{Right:} The aluminium enclosure with all internal detector assembly components installed. The SIPHRA CoB can be seen in the centre surrounded by one of the PTFE spacers which supports it.}
\label{fig:enclosurespacers}
\end{figure}

To fully constrain the scintillator and prevent it from moving during launch, it is securely held within the enclosure by four 2.5\,mm thick PTFE corner inserts which can be seen in Figure~\ref{fig:enclosurespacers}.
A 2\,mm thick PTFE insert is placed at the back of the scintillator which includes a serpentine venting channel leading to a 1\,mm hole in the end-cap at that end of the enclosure. This provides sufficient air evacuation capacity for the volume inside the enclosure while also preventing light ingress. 

The detector assembly electronic stack comprising the SiPM array, the SIPHRA ASIC, and the interface PCB are also supported within the aluminium enclosure by PTFE spacers which match the internal profile of the enclosure.
A series of spacers which are machined to match the contours and features of the PCBs are placed between each PCB in the stack and between the PCBs and the enclosure end-cap.
The harnesses from the interface PCB to the motherboard are routed along slots in the spacers and through cable channels machined into the end-cap.
The PCB spacers and the harness routing can be seen in Figures~\ref{fig:sipmarrays} \& \ref{fig:enclosurespacers}.

A silicone optical pad is used as an interface between the scintillator window and the SiPMs. These pads are suitable for use in a vacuum and are slightly deformable.
The dimensions of the PTFE spacers and the enclosure are designed such that as the end-caps are installed, the silicone pad is slightly compressed achieving good optical coupling and securing all components within the enclosure.

The PTFE spacers also serve to insulate the internal components from the enclosure.
Combined with the relatively large thermal mass of the scintillator, the spacers result in internal temperatures which are more stable against external fluctuations.
This functionality was demonstrated during thermal vacuum testing of the GMOD instrument \cite{Mangan2021}.
During temperature slews, the temperature of the internal components, measured at the SiPM array and at the aluminium side wall of the scintillator hermetic housing, was seen to significantly lag behind that of the enclosure which was in direct contact with the TVAC chamber's thermal plate.
In flight, the temperature fluctuations are expected to be even further attenuated as the temperature of the internal stack components will lag behind that of the spacecraft exterior.

%

\section{GMOD Motherboard}
\label{sec:gmodmotherboard}
The GMOD Motherboard is a CubeSat PC-104 compatible PCB which hosts the support electronics necessary to operate the detector assembly. It includes a micro-controller, responsible for data collection, and a CPLD for interfacing to SIPHRA's serial data output. An SPI flash memory allows the data to be cached in GMOD before being sent to the spacecraft on-board computer.  

The motherboard also hosts the power conditioning for the detector assembly including the adjustable bias PSU which generates the negative bias voltage required by the SIPM array.
Separate linear voltage regulators are provided for each of SIPHRA's power rails, resulting in a stable low-noise power supply.
The 80\,\si{\micro\ampere} constant reference current required by SIPHRA is provided by an adjustable constant current source based on a reference implementation from Texas Instruments \cite{slau507} with component values adjusted to match the desired output range.

\begin{figure}[h]
\includegraphics[width=0.9\textwidth]{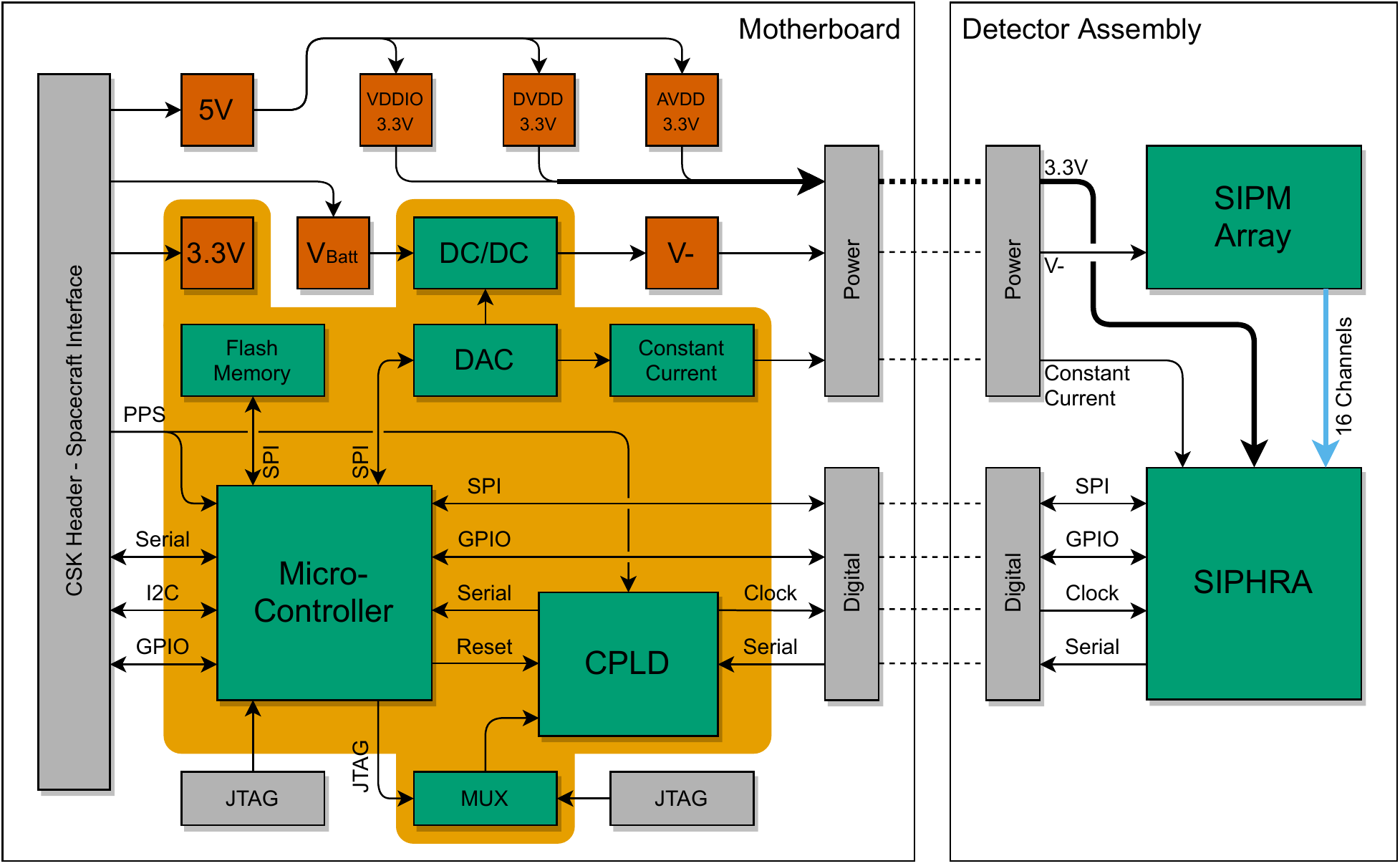}
\caption{Block diagram of the GMOD motherboard and detector assembly showing the connections between components. Green indicates a major functional component, orange indicates power supply nets and grey indicates a physical connector interface.}
\label{fig:gmodblock}
\end{figure}

A block diagram illustrating the functions of GMOD is shown in Figure~\ref{fig:gmodblock}.

\begin{figure}
\includegraphics[height=5.6cm]{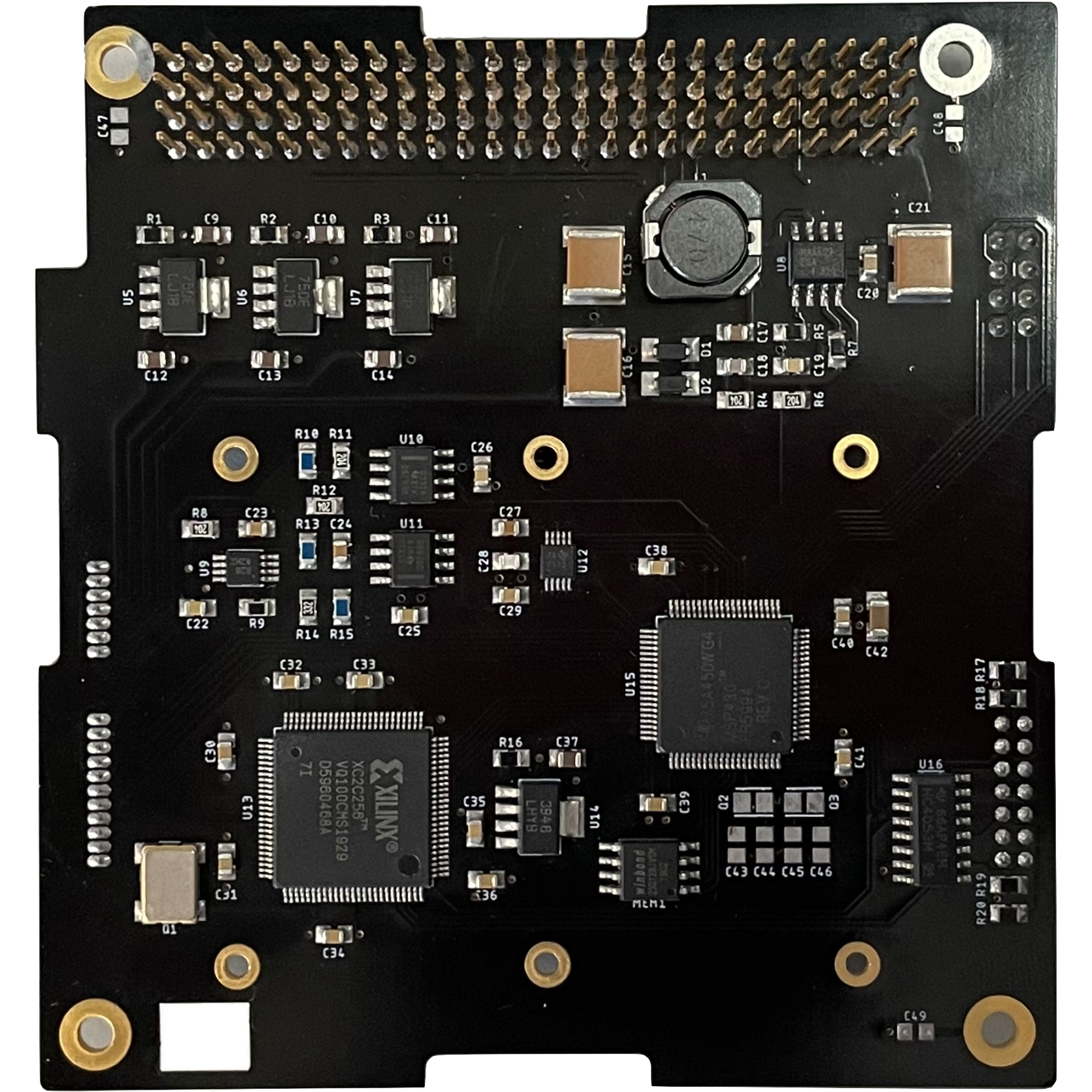}
\hspace{0.5cm}
\includegraphics[height=5.6cm]{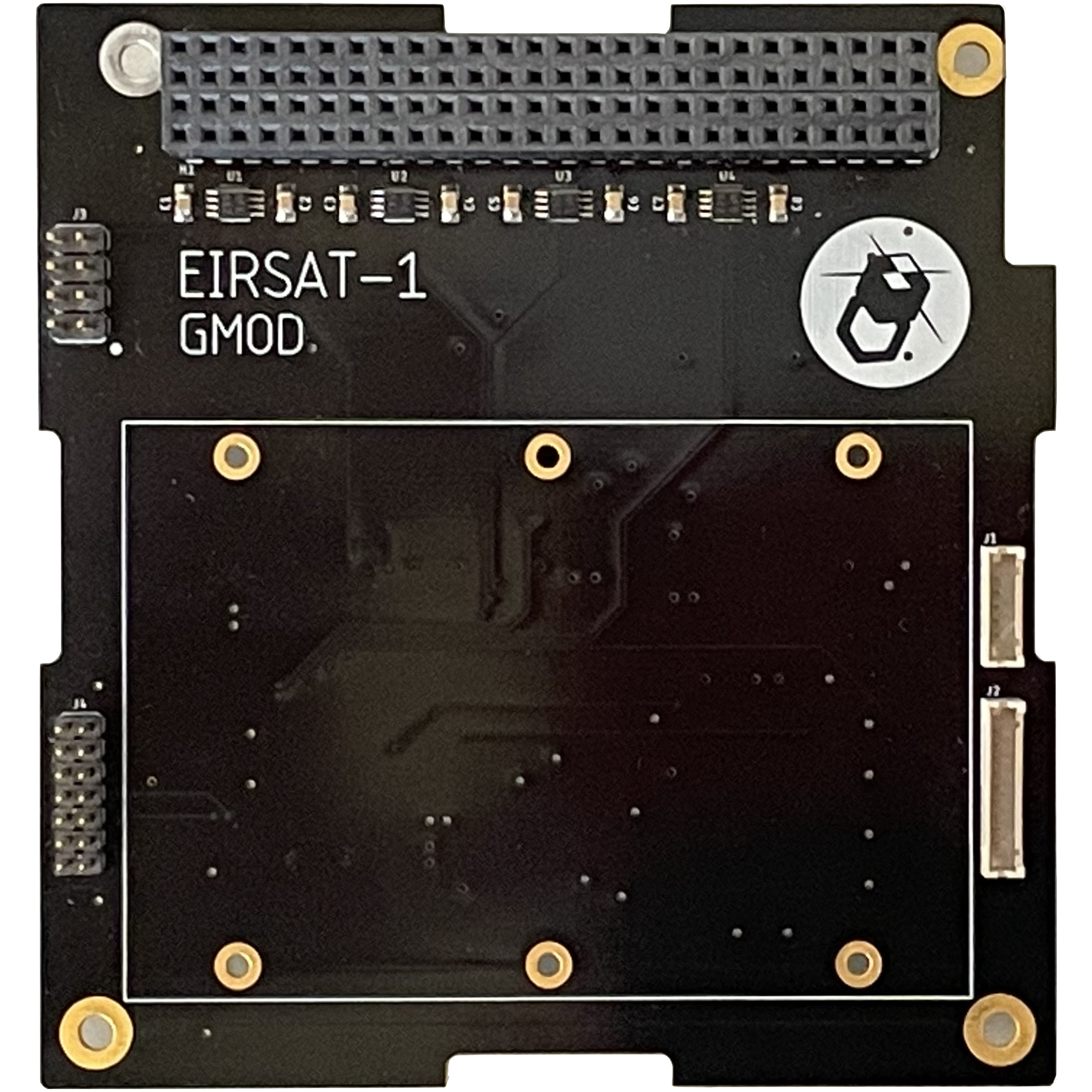}
\caption{The GMOD motherboard. The majority of the components are accommodated on the bottom of the board (left) allowing the detector assembly to be bolted to the top of the board (right).}
\label{fig:gmodmobopic}
\end{figure}

The motherboard conforms to the de facto standard for CubeSat components popularised by the Pumpkin Inc. CubeSat Kit, measuring 90.17\,mm\addthinspace$\times$\,95.89\,mm and interfacing to EIRSAT-1's main system bus through a pair of Samtec ESQ-126-39-G-D stack-through headers.
The motherboard is shown in Figure~\ref{fig:gmodmobopic}.

The board has been designed with the majority of its components on the underside thereby allowing the 75\,mm\,$\times$\,51\,mm detector assembly to be bolted directly to the top of the board.
The system bus interface header carries I$^2$C, serial, PPS, and other logic signals between GMOD and the OBC. Power from the Electrical Power System (EPS) is also provided to GMOD via this header. 
The motherboard has four other connectors, two of which are used for connection to the detector assembly as detailed in Section~\ref{sec:gmoddetector} and two JTAG connections which are used to reprogram the micro-controller and the CPLD.

%

\subsection{Bias PSU}
\label{sec:gmodbiaspsu}
The bias power supply unit (PSU) is designed to provide an adjustable bias voltage between $-25$\,V and $-28.3$\,V suitable for the SiPM array.

The bias PSU is based on the MAX629 DC-DC converter from Maxim Integrated, operated in an inverting boost mode, providing an output that is both negative and greater in magnitude than the input voltage.
The DC-DC converter is supplied by a switchable EPS channel at the battery voltage which is dedicated to GMOD.
The suitable range of the PSU is determined by the SiPM breakdown voltage and the desired SiPM gain which is related to the over-voltage above breakdown at which the SiPMs are biased.
The SensL J-series SiPMs have a breakdown voltage of approximately 24.5\,V and for GMOD, an appropriate gain is found at an over-voltage of approximately 3\,V, leading to a nominal SiPM bias voltage of $-27.5$\,V as the array is reverse biased.

An adjustable output is necessary to either maintain a consistent gain by compensating for calibration drift in the PSU and for the effect of temperature on the SiPM breakdown voltage which varies by 21.5\,mV/\textdegree\,C \cite{jseries}, or to adjust the gain if desired.
The output of the DC-DC converter is adjustable by biasing the DC-DC converter's feedback with a voltage generated by a Texas Instruments DAC7562T Digital-to-Analog Converter which allows for 1.2\,mV increments.

The output of the bias PSU is monitored by the micro-controller at two points using the 12-bit ADC channels of the micro-controller.
The first is the PSU feedback bias voltage at the DAC output which is monitored directly, while the second is the output of the PSU which is scaled through an amplifier to convert the PSU's output range of $-25$\,V to $-28.3$\,V to between 0\,V and 3.3\,V.

%

\subsection{Complex Programmable Logic Device}
\label{sec:gmodcpld}
The serial stream (Section~\ref{sec:frontend}) requires programmable logic to efficiently parse.
The serial output from SIPHRA is an unusual format featuring 20-bit words and inverted idle, start, and stop conditions when compared to a standard serial interface.
It is therefore not compatible with any standard peripheral interface available on a micro-controller. 

As the logic needed is relatively simple, a Xilinx CoolRunner XC2C256 Complex Programmable Logic Device (CPLD) was chosen as it provides the required functionality with low power consumption. GMOD utilises the CPLD to
 process the SIPHRA serial data before transmitting a simplified data stream to the micro-controller using a standard serial interface. The CPLD also generates the external 1\,MHz clock signal which operates the readout controller. 
 
The SIPHRA readout controller transmits digitised signals as a series of 20-bit words, one for each channel which is readout during a trigger.
The word format includes a 5-bit channel ID, 3 bits of trigger information, and the 12-bit ADC value for the channel.
The CPLD reads this serial data using custom receiver logic, parses the channel ID and packs the trigger information and the ADC value into two bytes.
The CPLD is also responsible for timestamping, determining a `finetime' value in microseconds which is disciplined by a PPS input from the OBC.
This timestamp is generated when channel 0, the first channel for each trigger, is received from SIPHRA.
The CPLD separates the individual SiPM channels (Ch1--16) from the array temperature (Ch0) and the summed channel (Ch17) and transmits them separately in two packet types to the micro-controller.
The packet types are prepended with a different 2-byte attached synchronisation marker (ASM) to allow the micro-controller to identify and appropriately processs their contents.

The individual, or `16-Channel' packet contains the ASM, timestamp, and the trigger flags and ADC values for the 16 individual SiPM channels.
The summed packet contains the ASM, timestamp, and the trigger flags for the summed SiPM channel and the array temperature.
The data volume can be reduced by disabling readout of channels 1--16 on SIPHRA which the CPLD will recognise and only transmit the summed packet to the micro-controller.

The CPLD includes a JTAG programming interface that is connected via a multiplexer to both the micro-controller and a header with a standard Xilinx pinout. This configuration allows the CPLD to be reprogrammed on the ground and in-flight.

%

\subsection{Micro-controller}
\label{sec:gmodmsp}
A Texas Instruments MSP430FR5994 micro-controller operates all peripheral functions of GMOD and serves as the interface between the payload and the  OBC.
This micro-controller was selected based on power and radiation-tolerance considerations.
The `FR' range of MSP430 micro-controllers use Ferroelectric Random Access Memory (FRAM) which stores data in the electric polarity of a PZT (lead zirconate titanate) film and is highly tolerant to radiation effects.

The micro-controller is connected to the OBC via the main system bus header.
The connections include an I$^2$C interface to a dedicated payload I$^2$C bus, a serial connection with bootstrap loader (BSL) support, and OBC GPIOs which are used to reset the micro-controller and invoke the BSL for reprogramming in flight.
A JTAG programming interface is provided on the board for on ground reprogramming.
The I$^2$C interface operates as a command interface through which the OBC controls GMOD while the serial interface is primarily used to bulk transfer data collected by GMOD to the OBC for in-flight processing, storage, and downlink to the ground \cite{Doyle2020}. 

Using the peripheral hardware on the motherboard, the micro-controller is responsible for data collection and management, control of SIPHRA and control of power components.

\subsubsection{Data Collection and Management}
The micro-controller accepts event data generated by SIPHRA which has been preprocessed and formatted by the CPLD.
The event data is asynchronous, being generated in response to gamma-ray detection events in the detector, and is received by the micro-controller using a standard serial receiver.

The micro-controller sorts the event data based on the ASM type which is attached by the CPLD and stores the events in 256-byte data units which are prepended with a coarse time-stamp corresponding to the mission-elapsed time.
The complete data units are then both transmitted to the OBC and stored in the GMOD SPI flash. The SPI flash implements a circular buffer which acts as a backup as well as a cache which allows the event data rate to temporarily exceed the GMOD to OBC data rate.
The OBC can either request specific data units from the SPI flash or command GMOD to transmit data units as soon as they are available.

\subsubsection{Control of SIPHRA}
While event data from SIPHRA is read using a serial UART via the CPLD, control and configuration of SIPHRA is achieved through an SPI interface which allows SIPHRA's configuration registers to be programmed. 
The register values are first loaded by the OBC into the micro-controller where they are buffered and then may be transferred to SIPHRA upon command.
SIPHRA implements a hardware parity check on these registers to indicate if the contents may have been corrupted by a single event upset (SEU).
The error output on SIPHRA is connected to a GPIO on the micro-controller allowing a pin change interrupt to immediately detect an SEU at which point the micro-controller can reload the correct register values from the buffer.

\subsubsection{Control of Power Components}
As detailed in Section~\ref{sec:gmodbiaspsu}, the bias voltage of the SiPM array is controlled and monitored by the micro-controller.
The control of the bias voltage is achieved via an SPI connected DAC which produces a voltage which influences the DC-DC converter's feedback input.
The monitoring of the bias voltage is through two feedback channels, one directly from the DAC and one which is scaled from the array voltage, into two of the micro-controller ADC inputs.

As the gain of the SiPM array varies with both temperature and bias voltage, the bias voltage can be adjusted in order to compensate for temperature variations.
The temperature of the array is read out by SIPHRA with each event.
Based on the temperature of the SiPM array, the voltage feedback and the desired detector gain, the micro-controller periodically calculates the correct DAC value to achieve the appropriate SiPM bias voltage.

The constant-current source on the motherboard which is used by SIPHRA as a reference current is also adjustable via voltage input.
The DAC used to adjust the bias voltage is a dual channel DAC and its second output is utilised to allow micro-controller control of the reference current allowing for calibration to be performed without adjusting component values.


\section{Data Handling}
\label{sec:gmoddata}
Data handling responsibilities for GMOD are shared between the motherboard \cite{Mangan2021b} and EIRSAT-1's OBC.
GMOD produces data in the form of time-tagged events (TTEs, Section~\ref{sec:gmodtte}) which are considered the canonical GMOD data-type from which science data should be generated on the ground.
EIRSAT-1 uses a UHF downlink operating at a 9600\,baud and while multiple ground stations are envisaged, the mission feasibility has been conducted assuming a single ground station.
This single station, depending on the orbit, limits contact with EIRSAT-1 to several communication windows per day, each of which are each on the order of 10\,minutes \cite{Marshall2021,Dunwoody2021}.
Over several contact windows, an average total daily contact time of approximately 29 mins is available which is sufficient to downlink 2\,MB of data.
During nominal operations, approximately 1.5\,MB of this data can be allocated to GMOD.
To reduce the bandwidth, the OBC performs on-board generation of secondary data products, including light-curves (Section~\ref{sec:GMODlightcurves}), spectra (Section~\ref{sec:GMODspectra}), and event-rate triggers (Section~\ref{sec:GMODtriggers}), all based on TTEs from the summed SIPHRA channel.
The total light-curve and spectral data utilisation are configuration dependent, but at a very conservative estimate will generate less than 450\,kB per day, giving a remaining daily data budget of $\sim$1\,MB for the downlink of TTEs.
Based on the predicted in-orbit background rate described in \cite{Murphy2021}, the data generation rate for TTEs from the summed SIPHRA channel is estimated to be approximately 30\,MB per day.
It is therefore possible to downlink many but not all generated TTEs for ground processing with a single station.
Since the housekeeping needs of the spacecraft and downlink of GMOD secondary data products are met by a single ground station, TTE downlink can be improved with the provision of additional stations.

For each ground station, the downlink windows are clustered, with several occurring each one orbit apart, followed by large gaps of approximately 18--20\,hours during which downlinking is not possible.
With multiple ground stations, the distribution of contact windows becomes more complicated and is dependent on the geographical location of the stations.
As it will be difficult to locate ground stations in such a way that large gaps in contact will be eliminated, the operations strategy has been planned to accommodate these gaps and clustered communication windows.
During the first window, the secondary data products will be downlinked first to determine which TTEs should be scheduled for downlink for the remainder of the communication time in the cluster.
At the start of each following communication window, new secondary data products which have been generated during the orbit will be downlinked in order to adjust the TTE downlink priority, for example if TTEs associated with a new trigger have become available.

During large communication window gaps, timely notification of triggers will still be communicated to the community by incorporating the most recent trigger information into EIRSAT-1's beacon which is broadcast from the spacecraft at regular intervals, as fast as once per 30\,s depending on battery levels.
The beacon format will be publicly available allowing it to be received and decoded on the ground using amateur radio equipment or inexpensive software defined radios.
A hosted server will allow received beacon data to be submitted and made publicly available.
It is anticipated that larger community initiatives such as SatNOGS \cite{satnogs} will help to collate and publish EIRSAT-1 beacon data.

The support of the amateur radio community will be sought to downlink additional EIRSAT-1 data.
While this is complicated by the need to manage the downlink using cryptographically signed uplink commands, the EIRSAT-1 ground segment has been planned to allow communication through trusted suitable third-party ground stations, including amateur stations.

\begin{figure}
\includegraphics[width=\textwidth]{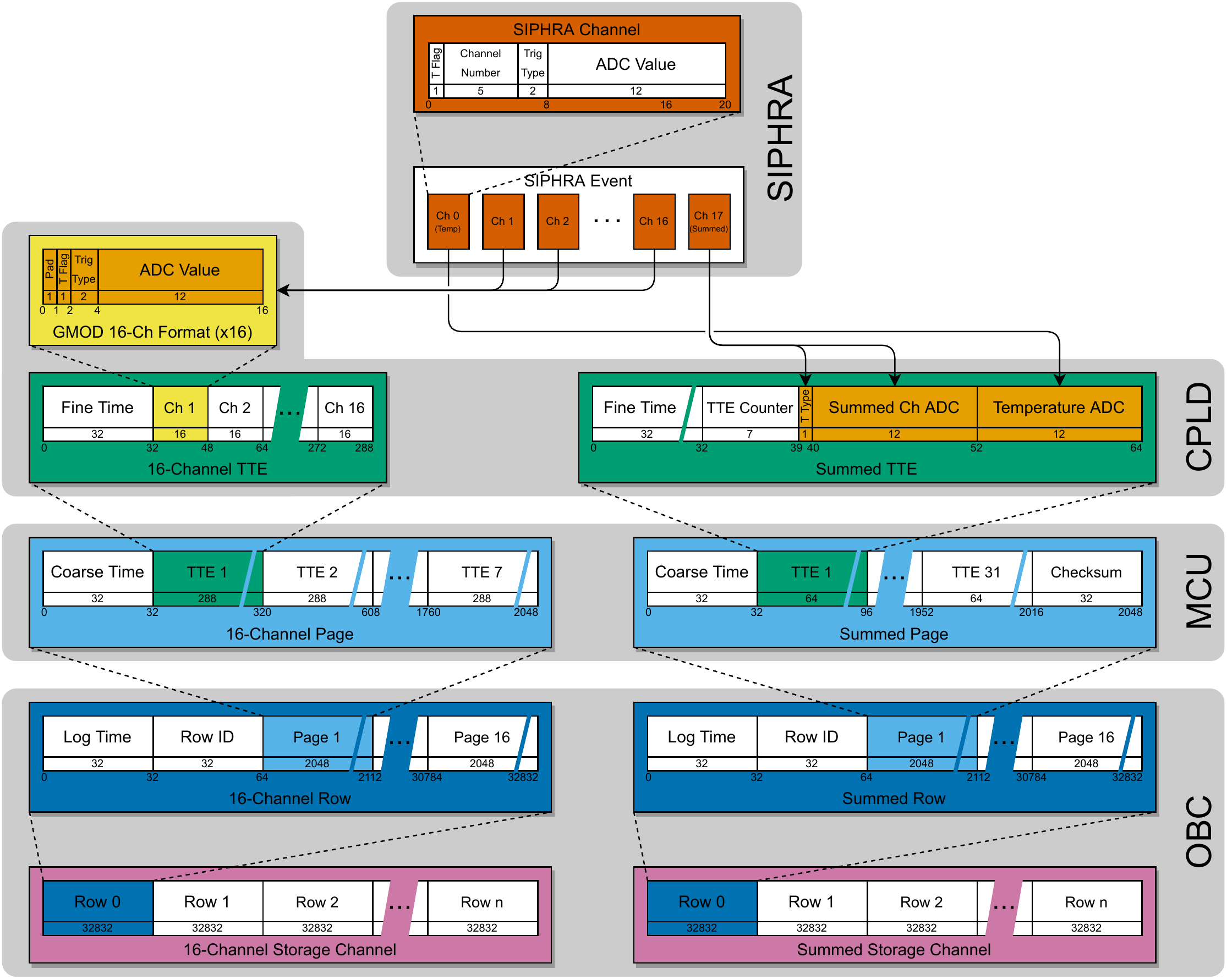}
\caption{The data-flow of time-tagged events indicating the data structures and formats used by each component in the processing chain. Events are generated by SIPHRA in the detector assembly, are time-tagged by a CPLD and processed by an MSP micro-controller on the GMOD motherboard before being transferred to EIRSAT-1's OBC for downlink to the ground and generation of secondary data products.}
\label{fig:gmoddata}
\end{figure}

\sloppy
Figure~\ref{fig:gmoddata} illustrates the data flow between GMOD components with a representation of the data structures utilised during each step.
GMOD data originates as event data in SIPHRA which is then fine-time-tagged and formatted by the CPLD, becoming separate `16-channel' and `summed-channel' TTEs before being transferred to the micro-controller.
The micro-controller sorts and packages multiple TTEs along with a 4-byte coarse-timestamp into 256-byte long data units referred to as `pages', a term inherited from the smallest writable data unit used by GMOD's flash memory.
Each summed-channel page holds 31 TTEs while a 16-channel page holds just 7 TTEs.
The coarse-timestamps are measured in seconds and are used to resolve the ambiguity of the TTE's finetime which overflows approximately every 71 minutes.
The 16-channel and summed-channel pages are then each cached by the micro-controller in both a circular buffer implemented in the flash memory and a FIFO buffer implemented in the micro-controller random access memory (RAM).
Pages are then transferred from GMOD to the OBC typically in an automatic `streaming' mode, but may also be manually requested.
In streaming mode, the OBC sets a ready flag when it has finished processing data, indicating to GMOD that it may transmit the next page when it becomes available.
In this mode, pages are typically taken from the RAM FIFO but when high event-rates occur this FIFO may not be large enough and the pages will instead be taken from the flash circular buffer until the backlog has been cleared.
Manually requested pages are retrieved based on their page address within the flash.

The pages received by the OBC are then stored in data storage channels on the OBC's error detecting and correcting flash memory.
Storage channels are a feature of OBC software \cite{Doyle2020} which is written using Bright Ascension's Flight Software Development Kit \cite{fsdk}. 
The storage channels are data structures which consist of a fixed number of rows, each of which have a fixed width.
Each row begins with the mission-elapsed-time at which it was saved and an incrementing absolute row ID.
As the pages are transferred to the OBC, the summed TTEs are processed to generate the light-curves, spectra, and triggers, which are also saved to their own data storage channels.

The OBC's storage channels may be either `circular' or `linear'.
Circular storage channels are used for most GMOD data-types and are written to one row at a time with the newest rows overwriting the oldest rows when the channel is full.
This is useful for storing the most recent data without requiring any data management.
Required data simply has to be downlinked to the ground before it is overwritten.
In contrast, linear storage channels are heavily write protected but require extensive data management.
Like circular channels, linear channels are also written to one row at a time but can no longer be written to when full.
Individual rows cannot be deleted so in order to free up space in a linear channel the contents of the entire channel must be deleted.
Linear channels are used to store a backup of TTEs associated with triggers (Section~\ref{sec:GMODtriggers}) to ensure that they are are not overwritten by newer data even if they cannot be downlinked in a timely manner.
Downlinking of all GMOD data-types to the ground is performed by requesting a range of rows from a particular storage channel which causes EIRSAT-1 to initiate a Large Data Transfer (LDT) compliant with the European Coordination on Space Standardization (ECSS) LDT service \cite{ECSS_E_ST_70_41C_2016}.

\subsection{Time-Tagged Events}
\label{sec:gmodtte}
As described in Section~\ref{sec:gmodasic}, for each gamma-ray event SIPHRA produces eighteen 12-bit ADC values encapsulated in eighteen 20-bit words also containing trigger flags and channel number.
The eighteen words correspond to array temperature (channel 0), SiPM channels 1--16, and the summed SiPM channel (channel 17) in that order.
Upon receipt of channel 0, the CPLD records a 4-byte fine-timestamp in microseconds which it then combines with trigger flags and ADC values from SIPHRA to produce two time-tagged event data-units with distinct formats, one for summed-channel data and one for 16-channel data.

The summed-channel TTEs are considered the most important data type as these will primarily be used for science while 16-channel TTEs are primarily for verification of the instrument's ability to perform individual SiPM readout, a feature which will be important in future detectors which require interaction localisation.
The summed-channel TTEs carry a sequential counter added by the CPLD to ensure that no events have been dropped from the micro-controller or OBC processing chain.
They are also more optimised for downlink by bit-packing, a process of concatenating the bits from several values which do not fully fill an integer number of bytes and then packing every 8 bits into a byte with no respect to the value boundaries, rather than padding each individual value to fill an integer number of bytes.
The summed-channel TTEs consist of the 4-byte fine-timestamp, a 7-bit sequence counter, a 1-bit readout flag indicating that the event was `internally' generated in SIPHRA by a gamma-ray event rather than externally generated by a forced readout, the 12-bit ADC value for the summed channel, and the 12-bit ADC value for the SiPM array temperature resulting in each summed-channel TTE being 8 bytes long.

The 16-channel processing is comparatively simpler, resulting in less space-efficient TTEs but which retain more triggering information which is useful for evaluation of the instrument.
Each 20-bit word corresponding to channels 1--16 has the 5 bits which indicate the channel number removed, resulting in 15 bits which are padded and packed into 2 bytes.
Channels are readout in a known order so the channel number is unnecessary.
The 15 bits consist of a 1-bit indicator that the individual channel was above readout threshold, a 2-bit indicator of readout type (internally generated, externally generated by the hold input, externally generated by SPI command), and the 12-bit ADC value for the channel.
The full 16-channel TTE contains the 4-byte fine-timestamp followed by the sixteen 2-byte channel values for a total of 36 bytes per 16-channel TTE.

The formats of both TTE types are summarised in Figure~\ref{fig:gmoddata}.


\subsection{Light-curves, Spectra, and Triggers}
\label{sec:GMODlcst}
Once the TTE data has been transferred to the OBC it is available for further on-board processing.
The GMOD component of the on-board software processes the summed TTE data to generate light-curves, spectra, and trigger data-types.
Once generated, all of the data-types are saved to respective storage channels allowing for downlink.

\subsubsection{Light-curves}
\label{sec:GMODlightcurves}
\begin{figure}[h]
\includegraphics[scale=0.75]{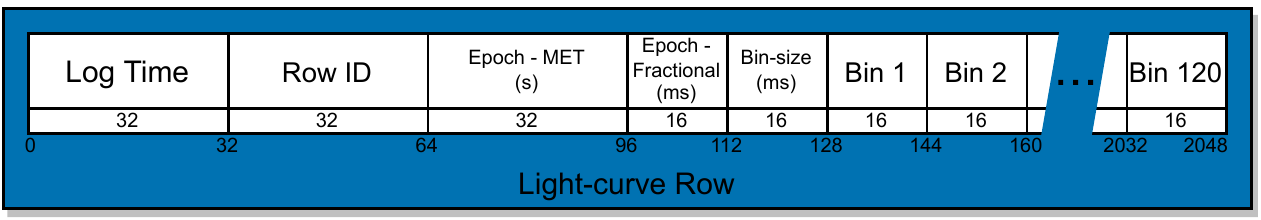}
\caption{GMOD light-curve data format.}
\label{fig:gmoddatalightcurve}
\end{figure}

The light-curve data-type records the number of counts over a configurable energy range in adjustable time bins.
The light-curve precision can be configured by setting the number of milliseconds for each bin.
The data-type consists of the light-curve epoch, bin duration, and uint16 values representing the number of counts in each bin.
The full light-curve row format is detailed in Figure~\ref{fig:gmoddatalightcurve}.

\subsubsection{Spectra}
\label{sec:GMODspectra}
\begin{figure}[h]
\includegraphics[scale=0.75]{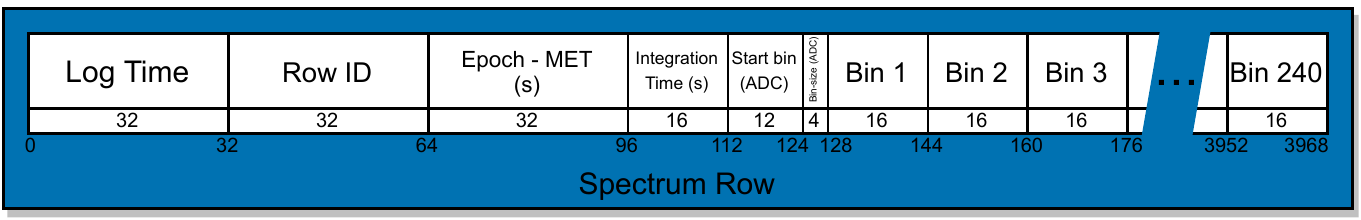}
\caption{GMOD spectrum data format.}
\label{fig:gmoddataspectrum}
\end{figure}

The spectrum data-type records the number of counts in discrete energy bins over an adjustable period of time.
GMOD produces TTEs containing 12-bit ADC values for photon energy giving 4096 possible energy bins.
The on-board generated spectrum data-type re-bins these values into adjustable energy bins beginning at a configurable ADC value and each containing a configurable integer number of ADC values between 1 and 16.
The data-type consists of the spectrum epoch, integration time, ADC value of the first bin, bin width, and uint16 values representing the number of counts in each bin.
The full spectrum row format is detailed in Figure~\ref{fig:gmoddataspectrum}.

\subsubsection{Triggers}
\label{sec:GMODtriggers}
One of the most important data-types for ascertaining interesting GMOD data is the trigger data.
Triggers are generated by calculating the ratio of two moving averages of counts within a certain energy range in order to estimate the current gamma-ray signal significance which is compared against a predefined significance trigger level.
The durations of the `signal' window, `background' window, and a background window delay can be tuned and reconfigured in-flight.
The intention is that the signal average represents the current gamma-ray flux and that the background average represents the current gamma-ray background.
The background window length is of the order of several tens of seconds and a delay of 10--20\,s excludes the most recently detected counts reducing an over-estimation of background for long GRBs.
Appropriate signal window lengths can vary for different sources so multiple trigger configurations will be processed in parallel with signal windows ranging from several tens of milli-seconds to several seconds.

In order to simplify the on-board algorithm for calculating the moving averages, they are not true instantaneous averages recalculated for the arrival of each gamma-ray photon but instead implemented as a series of time bins of short duration (i.e. 8\,ms) which are summed to reach the signal and background durations.

\begin{figure}[htb]
\includegraphics[scale=0.75]{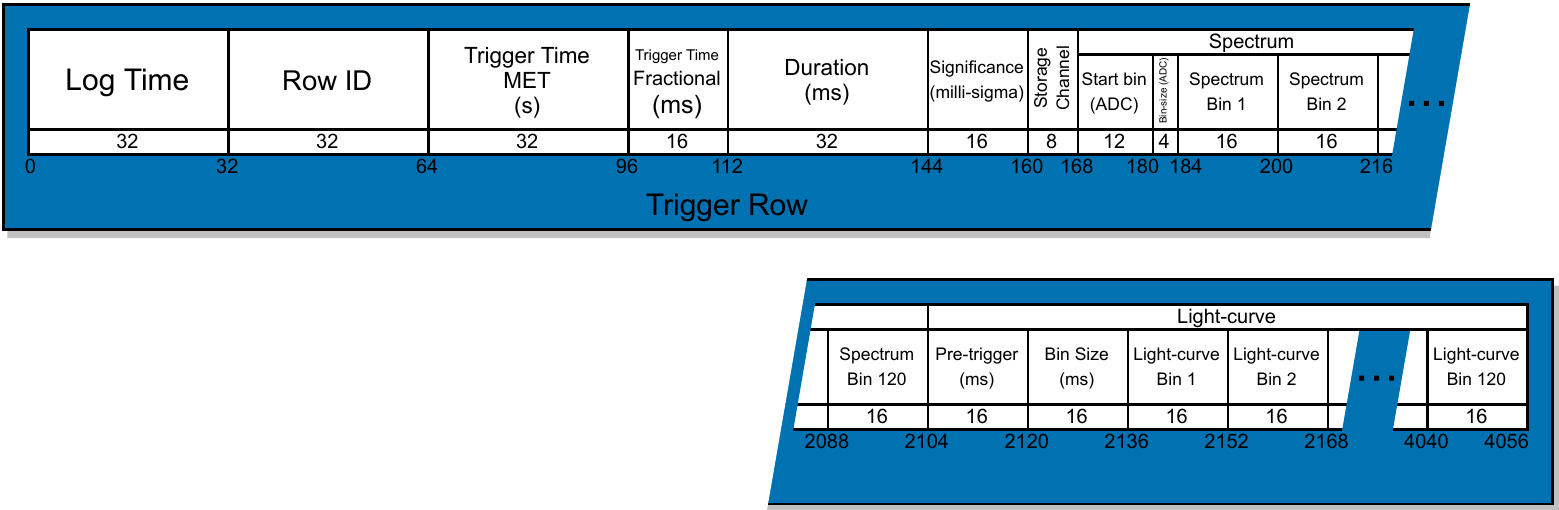}
\caption{GMOD trigger data format.}
\label{fig:gmoddatatrigger}
\end{figure}

When the significance level exceeds the trigger level, a number of important steps are carried out by the OBC.
Firstly, a trigger data-type is generated containing the trigger time, duration and peak significance of the burst, protected storage channel index, a trigger spectrum, and a trigger light-curve.
The full trigger format is detailed in Figure~\ref{fig:gmoddatatrigger}.
The protected storage channel index refers to a linear write-protected storage channel into which summed TTEs for a configurable time before and after the trigger are saved by the OBC.
Finally, the trigger data is incorporated into EIRSAT-1's beacon for immediate transmission to potential receivers on the ground.

\FloatBarrier

\section{Detector Resolution}
\label{sec:gmodchar}
\begin{figure}[htb]
\includegraphics[width=\textwidth]{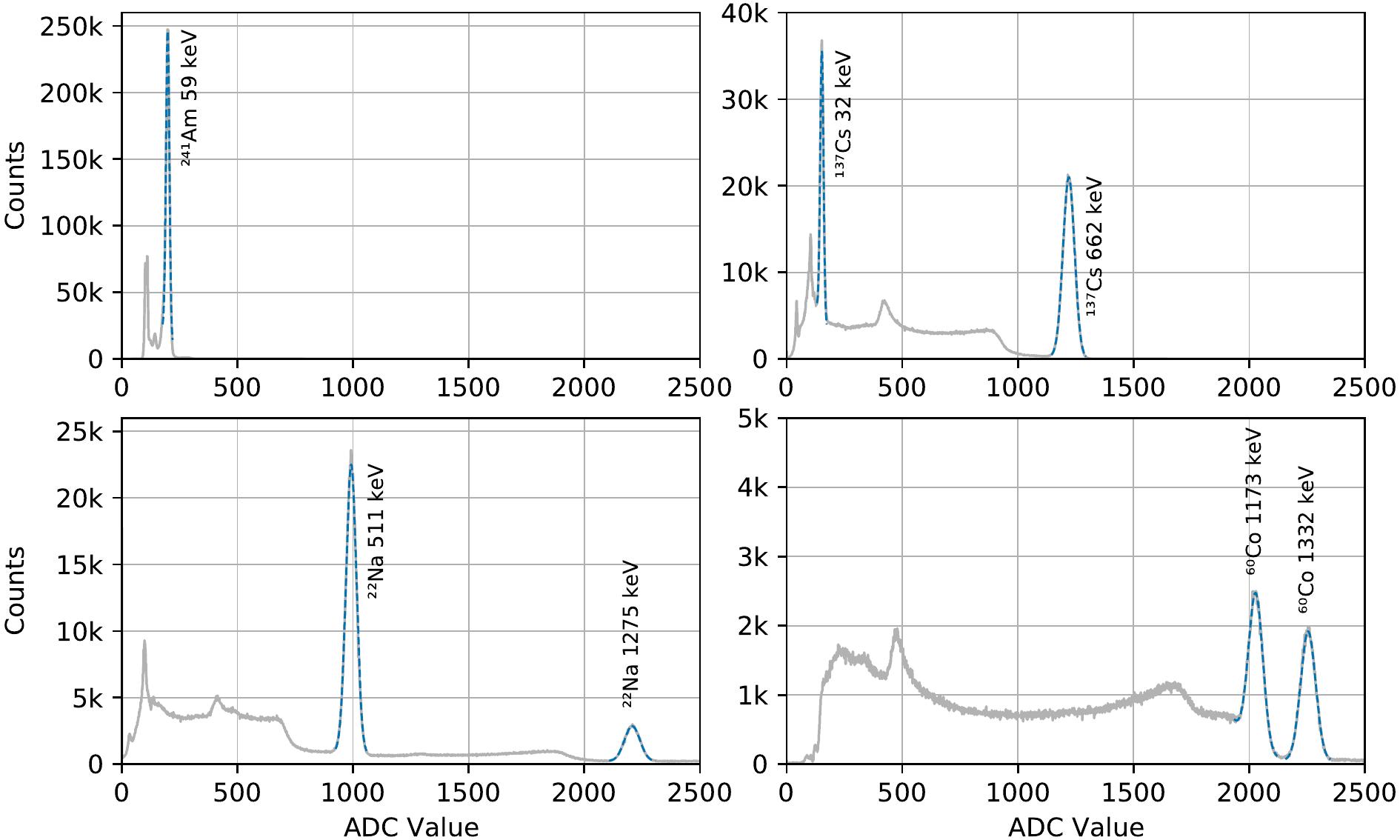}
\caption{Uncalibrated spectra of $^{241}$Am, $^{137}$Cs, $^{22}$Na, and $^{60}$Co measured using GMOD's summed-channel. The summed-channel combines the signals from all SiPMs prior to digitisation and is the primary data which will be used during in-flight operation. Several spectroscopic lines have been fit as indicated to determine a calibration for the detector in this mode.}
\label{fig:uncalspec}
\end{figure}

Following assembly of the engineering qualification model (EQM) of GMOD, four gamma-emitting radioactive nuclides were used to characterise the detector calibration and resolution: Americium-241, Caesium-137, Sodium-22 and Cobalt-60.
Measurements of the sources were made one at a time and sources were integrated for various durations depending on the source activity. 
This duration was typically around 30 minutes, though in the case of the weaker $^{60}$Co source the integration time was 90 minutes.
Figure~\ref{fig:uncalspec} shows the uncalibrated spectra of the four nuclides processed from summed-channel TTEs recorded by GMOD.
The spectra presented are essentially raw, with the downlinked data-set having been filtered only to ensure that all contributing events were internally generated by the GMOD detector assembly.
No re-binning has been applied and no baselines have been subtracted.

\begin{figure}[ht]
\includegraphics[width=\textwidth]{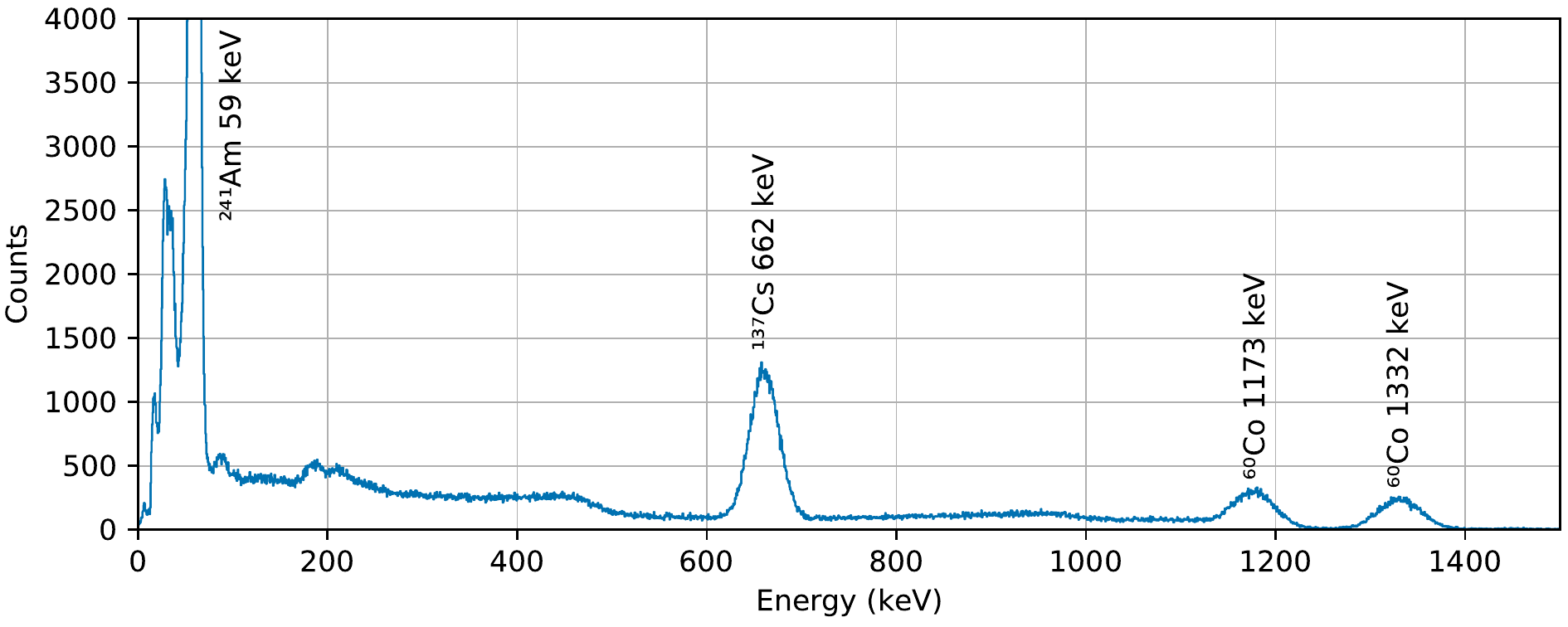}
\caption{Calibrated spectrum of a mixed test source containing $^{241}$Am, $^{137}$Cs, and $^{60}$Co measured using summed-channel data from GMOD.}
\label{fig:spec}
\end{figure}

The main spectral lines from each nuclide were fitted to determine their centre position.
The lines were fitted using a model consisting of a Gaussian curve to model the detector broadened mono-energetic emission line and a quadratic function to account for contributions from background and non-photo-peak measurements of higher energy lines.
A calibration for the detector was determined by fitting the known energy of the lines as a quadratic function of the measured centre position ADC value.
Figure~\ref{fig:spec} shows an example calibrated spectrum of a mixed test source containing $^{241}$Am, $^{137}$Cs, and $^{60}$Co processed based on this calibration from summed-channel TTEs recorded by GMOD.

Having calibrated the detector, the spectral lines from the four nuclides were refit in energy space to measure their widths and determine the detector spectral resolution.
The width of the Gaussian component of the fit gives the $1\sigma$ width of the line in energy, also called $\delta E$.
The full width at half maximum (FWHM) detector resolution for a given energy, $E$, is calculated as 2.35 times $\delta E$ expressed as a percentage of that energy.
The resolution of the detector at each of the measured spectral lines is given in Table~\ref{tab:resolution}.

\begin{table}
\caption{Spectral line widths as measured using summed-channel data from GMOD.}
\label{tab:resolution} 
\begin{tabular}{c S S S}
\hline\noalign{\smallskip}
\multicolumn{1}{l}{Nuclide} & \multicolumn{1}{l}{Energy (keV)} & \multicolumn{1}{l}{$\delta E$ (keV)} & \multicolumn{1}{l}{FWHM(\%)}  \\
\noalign{\smallskip}\hline\noalign{\smallskip}
$^{241}$Am &   59.541  &  4.0 & 15.9 \\
$^{137}$Cs &   32.061  &  4.1 & 29.7 \\
$^{137}$Cs &  661.657  & 15.3 &  5.4 \\
$^{22}$Na  &  511      & 13.5 &  6.2 \\
$^{22}$Na  & 1274.53   & 22.3 &  4.1 \\
$^{60}$Co  & 1173.237  & 20.6 &  4.1 \\
$^{60}$Co  & 1332.501  & 22.0 &  3.9 \\
\noalign{\smallskip}\hline
\end{tabular}
\end{table}

\begin{figure}[ht]
\includegraphics[width=0.8\textwidth]{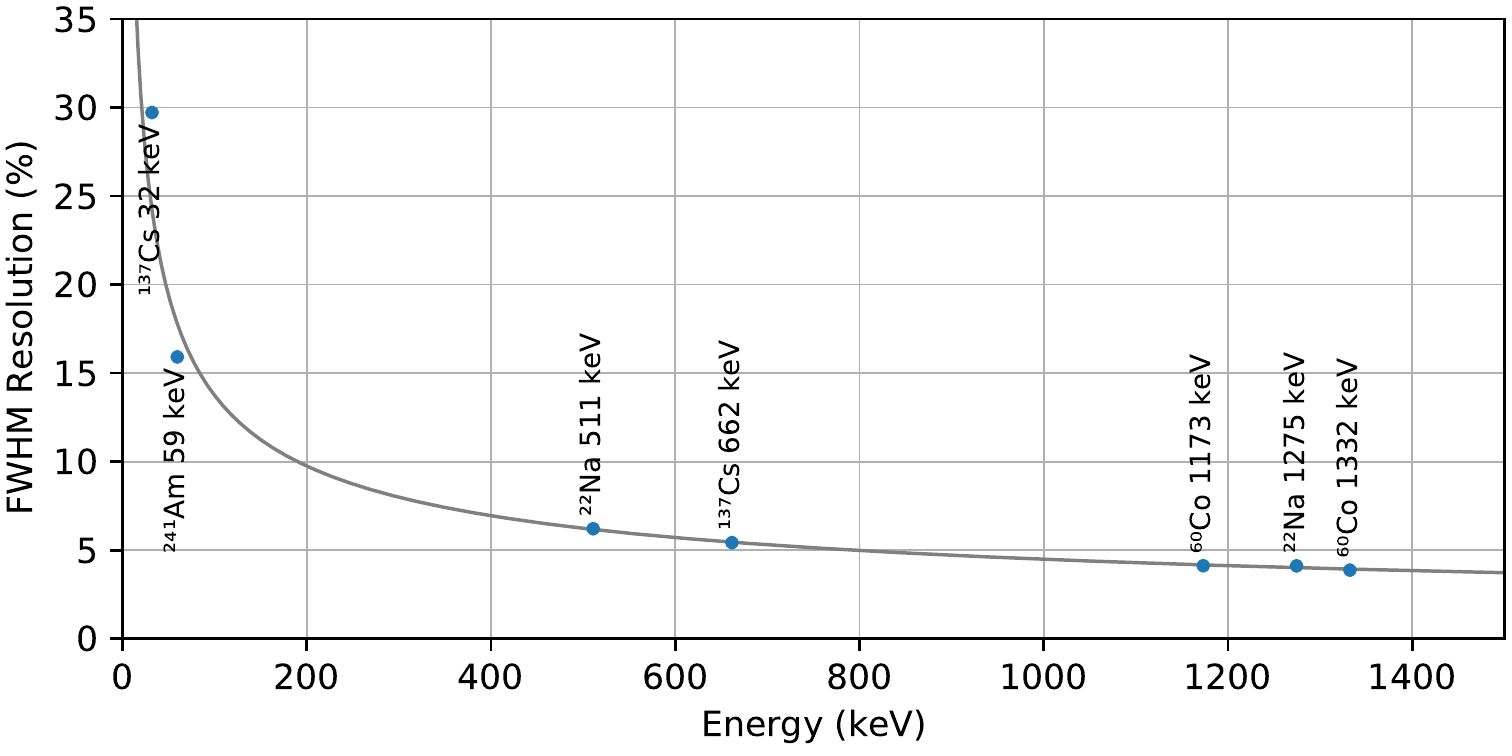}
\caption{Detector resolution as a function of energy. The labelled data points indicate the FWHM resolution as a percentage of measured energy based on measurements of spectral lines from radioactive nuclides while the solid line indicates the best fit of these data to Equation~\ref{eq:resolutionfit}.}
\label{fig:resolution}
\end{figure}

The values for the detector resolution were fit using the model
\begin{equation}
  \delta E = \sqrt{a^2 + b^2E + c^2E^2},
\label{eq:resolutionfit}
\end{equation}
which is adapted from \cite{bissaldi2009}.
The coefficient $a$ is a constant term representing limiting electronic resolution which is not expected to be an issue for scintillation detectors and is therefore set to zero in the fit.
The $b$ term is proportional to the square root of the energy, representing statistical fluctuations in the number of charge carriers which in turn is related to fluctuations in the number of scintillation photons or number of photoelectrons produced by the SiPMs.
The $c$ term is proportional to the energy, representing non-ideal transfer efficiency associated with the transfer of scintillation light to the SiPM. 
The fit coefficients were determined as $b = 0.583$ and $c = 5.033\times 10^{-3}$.
Based on this fit, Figure~\ref{fig:resolution} shows the detector FWHM resolution as a function of energy.

\section{Discussion and Conclusions}
\label{sec:gmoddiscussion}

GMOD is a miniaturised novel gamma-ray detector 
combining a CeBr$_{\rm3}$ scintillator, SiPMs and SIPHRA readout
in a CubeSat and  
 will be the primary payload of the EIRSAT-1.
This paper gives a comprehensive overview of its design, including its electronic and mechanical interfaces which are compatible with many off-the-shelf CubeSat systems and structures.
Table~\ref{tab:gmodparameters} summarises GMOD's key characteristics.

\begin{table}
\caption{GMOD key parameters.}
\label{tab:gmodparameters} 
\begin{tabular}{l l}
\hline\noalign{\smallskip}
Energy Resolution & 5.4\% ($@$662\,keV) \\
Effective Area            & 10\,cm$^2$ (sky-averaged) \cite{Murphy2021} \\
Energy Range              & 30\,keV--2\,MeV  \\
Detector Assembly Mass    & 336\,g  \\
Detector Assembly Volume  & 160\,cm$^3$ \\
Motherboard Mass          & 54\,g  \\
Motherboard Size          & 90.2\,mm\,$\times$\,95.9\,mm \\
Power Consumption         & $\approx$400\,mW \\
\noalign{\smallskip}\hline
\end{tabular}
\end{table}

GMOD demonstrates several promising technologies which have been identified as potential key components in the future generations of space-borne gamma-ray instruments, such as Compton or Compton-pair telescopes.
While several of these components have been demonstrated by other missions, the specific combination of scintillator, optical detector, and ASIC utilised by GMOD is unique.
Table~\ref{tab:missions} lists various missions and instruments either in development, or recently launched, utilising comparable technologies.
While CeBr$_3$ is currently in use as a scintillator paired with a photomultiplier tube on the BepiColombo MGNS instrument \cite{kozyrev2016}, GMOD will demonstrate its use with SiPMs in low Earth orbit.

\begin{table}
\begin{threeparttable}

\caption{Planned or recently launched instruments utilising emerging detector technologies on a CubeSat or SmallSat platform.}
\label{tab:missions}

\begin{tabular}{l c c c c c c}
\hline\noalign{\smallskip}
Mission / & \multirow{2}{*}{Scintillator} & Optical & \multirow{2}{*}{CubeSat} & Resolution & Effective & Energy\\
Instrument & & Sensor\tnote{1} & & $@$662\,keV & Area\tnote{2} & Range \\
\noalign{\smallskip}\hline\noalign{\smallskip}
EIRSAT-1/GMOD                       & CeBr$_3$     & SiPM  & 2U  & 5.4\%  & 10\,cm$^2$              & 30\,keV--2\,MeV  \\
HERMES \cite{Fuschino2020,Evangelista2020} & Ce:GAGG & SDD & 3U & $\approx$6.3\% & $\geq$50\,cm$^2$ & 2\,keV--2\,MeV  \\
GRBAlpha \cite{Werner2018,Pal2020}  & CsI(Tl)      & MPPC  & 1U  & -       & $\approx$56\,cm$^2$     & 10\,keV--300\,keV  \\
CAMELOT \cite{Werner2018,Ohno2018}  & CsI(Tl)      & MPPC  & 3U  & -       & $\approx$300\,cm$^2$    & 10\,keV--300\,keV  \\
BurstCube \cite{Perkins2020}        & CsI(Tl)      & MPPC  & 6U  & 8.6\%   & $\approx$70\,cm$^2$     & 50\,keV--1\,MeV  \\
GRID \cite{wen2019}                 & GAGG         & SiPM  & 6U  & 20\%    & $\approx$58\,cm$^2$     & 10\,keV--2\,MeV  \\
\noalign{\smallskip}\hline\noalign{\smallskip}
Glowbug \cite{Grove2020}            & CsI(Tl)/CLLB & SiPM  & No  & $<$4\%  & $\approx$1050\,cm$^2$  &  30\,keV--2\,MeV \\
SIRI-1 \cite{Mitchell2019}          & SrI$_2$      & SiPM  & No  & 4.3\%   &                        &  40\,keV--8\,MeV \\
GECAM \cite{Zhang2019}              & LaBr$_3$:Ce  & SiPM  & No  & 5.3\%   &                        &  6\,keV--5\,MeV  \\
\noalign{\smallskip}\hline
\end{tabular}

\begin{tablenotes}
\item[1] Though SiPM and MPPC may be used interchangeably, here SiPM denotes sensors manufactured by On Semiconductor (formerly SensL) while MPPC denotes those manufactured by Hammamatsu.
\item[2] The values quoted here are for the optimal incident angle. Many instruments, particularly those on CubeSats, feature a severe fall-off in area at non-optimal angles.
\end{tablenotes}

\end{threeparttable}
\end{table}

EIRSAT-1 is primarily an educational mission, being designed, built, tested, documented and operated by an interdisciplinary university student team. Its three novel payloads (including GMOD) incorporate technologies whose use in low-earth orbit will be demonstrated for the first time.
In addition, GMOD is predicted to detect $\sim$ 18 GRBs/year in the 50\,keV to 300\,keV energy range  \cite{Murphy2021}.
There is an ongoing global effort to rapidly communicate the detection of gamma-ray transients (e.g. as  counterparts to gravitational wave sources). To overcome the limited ground station coverage available, trigger results, including light-curves and spectra, will be created on-board and incorporated into the spacecraft beacon and transmitted continuously. Inexpensive hardware can be used to decode the beacon signal, making the data accessible to a wide community. 

A radiation damage study of GMOD's J-series SiPMs using 101.4 MeV protons has previously been carried out \cite{ulyanov2020}. The GMOD EQM has undergone environmental qualification at instrument level, qualifying the design for launch and low Earth orbit operation through multi-axis vibration and thermal-vacuum cycling tests \cite{Mangan2021}. The EIRSAT-1 spacecraft EQM  has been assembled, incorporating the GMOD EQM into the stack \cite{Walsh2020}. The completed EQM is undergoing functional testing \cite{Walsh2021} which will be followed by extensive mission testing \cite{Doyle2021} and a spacecraft level environmental test campaign.
Further detailed characterisation of the GMOD instrument, including determination of the off-axis response of the detector within the stack, will be performed on the EQM and used to verify the simulated performance estimates \cite{Murphy2021}.
The flight models of both GMOD and the EIRSAT-1 spacecraft will then be assembled and qualified to acceptance levels before delivery to ESA for subsequent launch.

GMOD will provide flight heritage for two other instruments for larger CubeSats, GIFTS and ComCube, both of which are based partly on the design of GMOD. ComCube is a Compton telescope concept currently in development, with several GMOD-like detector modules functioning as the D2 layer, and benefitting particularly from GMOD's pixellated SiPM array with individual SiPM readout, which allows localisation of gamma-ray interaction points within the monolithic scintillator \cite{ulyanov2017}.

\begin{acknowledgements}
The EIRSAT-1 project is carried out with the support of ESA’s Education Office under the Fly Your Satellite!\,2 programme. This study was supported by The European Space Agency's Science Programme under contract 4000104771/11/NL/CBi. JM, AU, DM, and SMB acknowledge support from Science Foundation Ireland (SFI) under grant number 17/CDA/4723. LH acknowledges support from SFI under grant 19/FFP/6777 and the EU AHEAD2020 project (grant agreement 871158). DM, RD, MD, CO'T and JT acknowledge support from the Irish Research Council (IRC) under grants GOIPG/2014/453, GOIPG/2019/2033, GOIPG/2018/2564, GOIPG/2017/1031, GOIPG/2014/685. SW acknowledges support from the European Space Agency under PRODEX contract number 400012071. JR acknowledges a scholarship from the UCD School of Physics. We acknowledge all students who have contributed to EIRSAT-1.
\end{acknowledgements}

\section{Data Availability}

The datasets generated during and/or analysed during the current study are available from the corresponding author on reasonable request.

\section{Conflict of Interest}
The authors declare no conflict of interest.

\bibliographystyle{spphys}       
\bibliography{refs}   

\end{document}